\documentclass[11pt,a4paper]{article}
\pdfoutput=1
\usepackage{jcappub}

\usepackage{xcolor}
\usepackage{bm}
\usepackage{subfigure}
\usepackage{feynmp}
\usepackage{extarrows}
\usepackage{slashed}
\usepackage{dcolumn}
\usepackage{verbatim}
\usepackage{here}
\usepackage{multirow} 
\allowdisplaybreaks[4]

\usepackage{ulem}

\numberwithin{equation}{section}
\renewcommand{\thefootnote}{\arabic{footnote}}

\newcommand{\fr}[2]{\mbox{$\frac{\,{#1}\,}{#2}$}}
\newcommand{\order}[1]{\mathcal{O}({#1})}

\def\bge{\begin{equation}}
\def\ede{\end{equation}}
\def\bga{\begin{aligned}}
\def\eda{\end{aligned}}
\def\bgp{\begin{pmatrix}}
\def\edp{\end{pmatrix}}
\def\bgs{\begin{subequations}}
\def\eds{\end{subequations}}
\newcommand{\beq}{\begin{equation}}
\newcommand{\eeq}{\end{equation}}
\newcommand{\bq}{\begin{equation}}
\newcommand{\eq}{\end{equation}}
\newcommand{\ba}{\begin{array}}
\newcommand{\ea}{\end{array}}
\newcommand{\beqa}{\begin{eqnarray}}
\newcommand{\eeqa}{\end{eqnarray}}
\newcommand{\beqs}{\begin{subequations}}
\newcommand{\eeqs}{\end{subequations}}

\def\dis{\displaystyle}
\def\di{{\mathrm{d}}}

\def\leqq{\leqslant}
\def\geqq{\geqslant}
\def\[{\left[}
\def\]{\right]}
\def\({\left(}
\def\){\right)}

\def\AA{\mathbf{A}}
\def\OO{\mathcal{O}}

\def\pd{\partial}

\def\to{\rightarrow}

\def\ii{\mathrm{i}}

\def\ga{\gamma}

\def\si{\sigma}

\def\ucut{\Lambda_{\text{U}}^{}}
\def\ucutt{\Lambda_{\text{U}}^2}

\def\Mp{M_{\mathrm{Pl}}}

\def\R{\mathcal{R}}
\def\X{\chi_s^{}}
\def\XX{\chi_s^2}
\def\MX{M_{\chi}^{}}

\def\xis{\xi_s^{}}

\def\xih{\xi_h^{}}
\def\ZZ{\mathbb{Z}_2^{}}
\def\vew{v_{\mathrm{\scriptscriptstyle{EW}}}}
\def\xfo{x_{\mathrm{f}}}
\def\fo{_{\mathrm{f}}}

\def\End{\end{document}}

\title{\huge Probing Gravitational Dark Matter}

\author[a,b]{\large Jing Ren,}
\author[a,c,d]{\large~ Hong-Jian He\,}

\affiliation[a]{~Institute of Modern Physics and Center for High Energy Physics,\\
                \hspace*{0.15mm} Tsinghua University, Beijing 100084, China}
\affiliation[b]{~Department of Physics, University of Toronto,
                 Toronto ON Canada M5S1A7}
\affiliation[c]{~Center for High Energy Physics, Peking University, Beijing 100871, China}
\affiliation[d]{~Kavli Institute for Theoretical Physics China, CAS, Beijing 100190, China}

\emailAdd{jingren2004@gmail.com, hjhe@tsinghua.edu.cn}

\abstract{
\\
 So far all evidences of dark matter (DM) come from astrophysical and cosmological observations,
 due to gravitational interactions of the DM. It is possible that the true DM particle in the
 universe joins gravitational interactions only, but nothing else. Such a Gravitational DM (GDM)
 may act as a weakly interacting massive particle (WIMP), which is conceptually simple and attractive.
 In this work, we explore this direction by constructing the simplest scalar GDM particle
 $\,\chi_s^{}\,$.\, It is a $\,\mathbb{Z}_2^{}$ odd singlet under the standard model (SM) gauge group,
 and naturally joins the unique dimension-4 interaction with Ricci curvature, $\,\xi_s^{}\XX {\cal R}\,$,\,
 where $\,\xi_s^{}$ is the dimensionless nonminimal coupling. We demonstrate that this gravitational
 interaction $\,\xi_s^{}\XX {\cal R}\,$,\, together with Higgs-curvature nonminimal coupling term
 $\,\xi_h^{}H^\dag H {\cal R}\,$,\, induces effective couplings between $\,\XX\,$ and SM fields,
 and can account for the observed DM thermal relic abundance. We analyze the annihilation cross
 sections of GDM particles and derive the viable parameter space for realizing the DM thermal relic
 density. We further study the direct/indirect detections and the collider signatures of such a
 scalar GDM. These turn out to be highly predictive and testable.
}

\keywords{Dark Matter, Quantum Gravity Phenomenology
\\[5mm]
JCAP (2015), Final Version [\,arXiv:1410.6436\,].
}

\begin{document}

\maketitle


\setlength{\baselineskip}{18pt}
\setcounter{page}{2}
\setcounter{footnote}{0}
\setcounter{figure}{0}
\renewcommand{\thefootnote}{\arabic{footnote}}

\vspace*{10mm}

\section{\hspace*{-3.5mm} Introduction}
\label{sec:1}
\vspace*{2mm}

All evidences of dark matter (DM) come from astrophysical and cosmological observations so far,
due to gravitational interactions of the DM. It is possible that Nature may have designed
the DM particle to join gravitational interactions only, but nothing else.
Such a Gravitational DM (GDM) acts as a weakly interacting massive particle (WIMP),
which is conceptually simple and attractive.

\vspace*{1mm}

The standard model (SM) of particle physics successfully describes
the electromagnetic, weak and strong forces in nature,
while the gravitation is best theorized by Einstein general relativity (GR).
It is apparent that the world is described by the joint effective theory\,\cite{EFT}
of the SM and GR, which could be valid up to high scales below the Planck mass.
We are well motivated to study the intersection between the SM and GR within this effective theory.
In this work, we construct the simplest scalar GDM particle $\,\X\,$,\,
which is a $\ZZ$-odd singlet under the SM gauge group,
and joins gravitational interaction only.
As such, there is a unique dimension-4 operator prescribing the interaction between
the GDM $\,\X\,$ and the Ricci curvature $\,\R\,$,\,

\begin{eqnarray}
\label{eq:NMC}
S_{\textrm{NMC}}^{} ~=\int\!\! d^4x\,\frac{~\xis\,}{2}\chi_s^2\,\mathcal{R}\, ,
\end{eqnarray}
where $\,\xis\,$ is the corresponding dimensionless nonminimal coupling.
Since all SM particles enjoy gravitational interaction,
gravity can serve as the natural messenger between the GDM and SM particles via \eqref{eq:NMC}.
In the present work, we systematically study the constraints and tests of such a GDM
for a variety of dark matter phenomenologies.
In passing, we also note that a recent different study considered a gravity-mediated
(composite) dark matter model in the context of warped extra-dimensions,
where the radion and massive KK gravitons serve as the mediator \cite{ExdGDM}.

\vspace*{1mm}

This paper is organized as follows.
In Section\,\ref{sec:2}, we present the minimal construction of GDM in
both Jordan and Einstein frames. Then, in Section\,\ref{sec:3},
we analyze the GDM as a WIMP dark matter candidate
and identify its viable parameter space for generating the observed dark matter relic abundance.
Section\,\ref{sec:4} is devoted to the systematical analysis of (in)direct searches of
the GDM, and the probe of the GDM at high energy hadron colliders.
We finally conclude in Section\,\ref{sec:5}.
Appendix\,\ref{app:loopfactor} will present
the formulas of radiative loop factors as needed for the physical applications
in Sections\,\ref{sec:3}-\ref{sec:4}. In Appendix\,\ref{app:TRDdetail}, we
calculate the threshold and resonance effects for dark matter annihilations,
which are needed for the thermal relic density analysis in Section\,\ref{sec:3}.

\vspace*{4mm}
\section{\hspace*{-3.5mm} Minimal Gravitational Dark Matter}
\label{sec:2}
\vspace*{2mm}

In this section, we present the formulation of the scalar GDM $\,\X\,$ and derive its induced interactions
with the SM particles. We first consider the GDM in Jordan frame, where the nonminimal coupling (\ref{eq:NMC})
is manifest. Then, we make the Weyl transformation on the metric and convert the action into Einstein frame,
in which the nonminimal term (\ref{eq:NMC}) is fully transformed away and result in a new set of effective
operators.  With these, we will systematically derive the relevant Feynman vertices for $\,\X\,$
in Einstein frame.

\vspace*{3.5mm}
\subsection{\hspace*{-3.5mm} Minimal GDM in Jordan Frame}
\label{sec:2.1}
\vspace*{2mm}

Within the joint effective theory of the SM\,+\,GR, we can write down the effective action by
including this scalar GDM field $\,\X\,$,
\begin{eqnarray}
\label{SRNMCJF}
S_{\text{J}}^{} &=& \!\int\!\! d^4x \mbox{$\sqrt{-g^{(J)}}$}
\bigg[\,\frac{1}{2}M^2\mathcal{R}^{(J)}_{}-\frac{1}{4}F^a_{j\mu\nu}F^{a\mu\nu}_j
+(D_\mu H)^\dag (D^\mu H) + \frac{1}{2}\partial_\mu\X\partial^\mu\X - V(H,\X)~~~~
\nonumber \\
&& \hspace*{25mm}
   +\frac{\,\xis\,}{2}\chi_s^2\,\mathcal{R}^{(J)}+\xih H^\dag H \mathcal{R}^{(J)}
   + {\cal L}_{\text{F}}^{}\,
\bigg] ,
\end{eqnarray}
where $\,g_{\mu\nu}^{(J)}\,$ and $\,\mathcal{R}^{(J)}_{}\,$
denote the Jordan frame metric and Ricci scalar, respectively.\footnote{%
We note that Ref.\,\cite{Xinflaton} considered a real scalar serving as both the DM particle and the inflaton
in the early universe. With $\xi_s^{}\gg\xi_h^{}$,\, inflation occurs along the real scalar direction.
To account for the cosmic fluctuation strength, the demanded $\,\xi_s^{}$ is far below $10^{15}$,\,
and thus fully differs from the relevant parameter range of $\,\xi_s^{}$ in the present paper
(cf.\ our Fig.\,\ref{fig:3}).
The nonminimal coupling in Ref.\,\cite{Xinflaton} is negligible for low energy DM phenomenology,
and its DM interacts with SM particles mainly via the conventional Higgs portal coupling $\,\lambda_{h\chi}^{}\,$
[cf.\ \eqref{eq:V}]. Hence, our current GDM construction realizes
a different DM mechanism from Ref.\,\cite{Xinflaton}
and invokes different parameter space of the DM nonminimal coupling.}\,
The Lagrangian term $\,{\cal L}_{\text{F}}^{}\,$ represents the fermion sector of the SM.
In Eq.\,\eqref{SRNMCJF}, we define the gauge field strength,
$\,F^a_{j\mu\nu}=(G^a_{\mu\nu},\, W^a_{\mu\nu},\, B_{\mu\nu}^{})$,\,
as well as the Higgs doublet field,
$\,H=\big(\pi^+\!,\,\frac{1}{\sqrt{2}}(\vew+\hat\phi+i\pi^0)\big)^T$,\,
where $\,\vew \simeq 246\,$GeV is the vacuum expectation value (VEV) of the SM Higgs at electroweak vacuum.\,
The second line of \eqref{SRNMCJF} contains the nonminimal coupling terms for the GDM $\,\X\,$
and the Higgs doublet $\,H\,$.\,
According to the constraints from the current LHC Higgs data\,\cite{atkins}
and from the perturbative unitarity\,\cite{XRH}, the nonminimal coupling $\,\xih\,$ receives
an upper limit around $\,\mathcal{O}(10^{15})$.\,
In the electroweak vacuum, the Higgs non-minimal coupling term makes a contribution to
the Einstein-Hilbert action:
$\,\frac{1}{2}M^2\mathcal{R}^{(J)}\to\frac{1}{2}(M^2+\xih \vew^2)\mathcal{R}^{(J)}\,$.\,
Hence, we can identify $\,M^2+\xih \vew^2=\Mp^2\,$,\,
where $\,\Mp = (8\pi G)^{-1/2} \simeq 2.44\times 10^{18}\,$GeV is the reduced Planck mass.
Given the existing constraint $\,\xih \lesssim \order{10^{15}}\,$,\,
we have $\,\xih \vew^2\ll \Mp^2\,$,\, and thus $\,M \simeq \Mp^{}\,$ holds to good accuracy.
We may also add the higher curvature terms $\,R^2$\, and $\,R_{\mu\nu}^{}R^{\mu\nu}$\, to the effective
action \eqref{SRNMCJF} as well. But they do not affect the leading-order graviton contributions,
and are irrelevant to the present analysis of scalar GDM.

\vspace*{1mm}

In Eq.\,\eqref{SRNMCJF}, $\,V(H,\chi_s)\,$ is the general potential including
the SM Higgs doublet $\,H$\, and the scalar GDM $\,\X\,$.\,
We construct $\,\X\,$ as a $\ZZ$-odd real singlet, which has vanishing VEV,
$\,\langle\X\rangle = 0$\,.\,
Then, we deduce the gauge-invariant scalar potential with CP and $\,\ZZ$\, symmetries
as follows,
\begin{eqnarray}
\label{eq:V}
V(H,\chi_s) ~=~
\lambda_h^{}\!\left(\! H^\dag H-\frac{\vew^2}{2}\right)^{\!\!2}
+\frac{\lambda_{h\chi}^{}}{2}\!\left(H^\dag H-\frac{\vew^2}{2}\right)\!\chi_s^2
+\frac{1}{2}M_{\X}^2\XX
+\frac{\lambda_\chi}{4!}\chi_s^4\, .~~~~
\hspace*{6mm}
\end{eqnarray}
Since we consider that the GDM field $\,\X\,$ joins gravitational interactions only,
the $\,\X\,$ has no direct coupling with the SM particles except coupling to gravity
and itself.  So we will set the Higgs portal coupling $\,\lambda_{h\chi}^{}\!=0\,$,\,
or be negligible for the current study.
Note that if $\,\lambda_{h\chi}^{}\!=0\,$ holds at tree-level, we might expect it to be
reinduced via nonminimal couplings $(\xih,\,\xis)$ due to graviton-exchange.
But such graviton-exchanges just induce a new dimension-6 effective operator
$\,(H^\dag H)\partial^2\XX\,$ in Einstein frame
[cf.\ Eq.\,\eqref{SRNMCEFb}],\footnote{%
This new dimension-6 operator $\,(H^\dag H)\partial^2\XX\,$
will play an important role for our analysis of the thermal relic density of GDM,
as shown in Eq.\,\eqref{Lssint} and Fig.\,\ref{fig:2} of Sec.\,\ref{sec:3}.}\,
which differs from the $\lambda_{h\chi}^{}$ term of dimension-4.\,
We also note that the graviton-loop may induce Higgs portal term. This should have a
coefficient proportional to $\,\xis\xih\Lambda^4/\Mp^4$,\,
where $\,\Lambda$\, is the UV cutoff for loop integration.
Setting $\,\Lambda$\, as the unitarity bound
$\,\Lambda\sim \Mp^{}/\sqrt{|\xis\xih|}\,$ (cf.\ Sec.\,\ref{sec:2.3}),
we can estimate the graviton-loop-induced Higgs portal coupling
$\,\lambda_{h\chi}^{}\propto |\xis\xih|^{-1}\ll 1\,$,\,
which is negligible for $\,|\xis\xih|\gg 1\,$ in the present study (cf.\ Sec.\,\ref{sec:3}).
In practice, we only need to mildly set $\,\lambda_{h\chi}^{}\lesssim \OO(10^{-2})\,$
for our construction, which has negligible contribution to the DM thermal relic density.
We also note that the Higgs portal term  $\,\lambda_{h\chi}^{}H^\dag H\XX\,$
was extensively studied in the literature\,\cite{HP} for realizing $\X$ as a DM.
It induces interactions of DM with other SM particles
and may provide the DM relic density if the coupling is sizable, \,$\lambda_{h\chi}^{}=\OO (0.1-1)$.\,
In an extended scheme, we may consider both couplings
$\,\lambda_{h\chi}^{}\,$ and $\,(\xis ,\,\xih)\,$
to give comparable contributions to the DM relic density.
But we will focus on the minimal GDM construction for the present study,
where the DM interacts with SM particles only via gravity-induced interactions.

\vspace*{1mm}

From the Jordan frame action (\ref{SRNMCJF}), the dominant interactions for the GDM arise from
its nonminimal coupling with the Ricci curvature.
We perturb the metric under flat background,
$\,g_{\mu\nu}^{(J)}=\eta_{\mu\nu}^{}+\kappa \hat{h}_{\mu\nu}^{}$,\,
where $\,\kappa\equiv\sqrt{2}/\Mp$\, and $\,\hat{h}_{\mu\nu}^{}\,$ denotes graviton.
Since $\,\langle\X\rangle=0$\,,\,
there is no mixing between $\,\hat{h}_{\mu\nu}^{}\,$ and $\,\X\,$.\,
Then, we derive the Feynman vertex for gravity induced triple coupling
$\,\X (p_1^{})\!-\!\X (p_2^{})\!-\!{h}_{\mu\nu}^{}(p)$\,,\footnote{%
With nonzero coupling $\,\xi_h^{}\,$ and Higgs VEV,
there is a kinetic mixing between the metric fluctuation $\,\hat{h}_{\mu\nu}^{}\,$
and the Higgs field $\,\hat\phi\,$.\,  After kinetic diagonalization,
$\,\hat\phi\,$ is rescaled as $\,\hat\phi =\zeta\phi\,$,\,
and the canonical Higgs field is the same $\,\phi\,$
as we will derive in Einstein frame [cf.\ Eqs.\,\eqref{eq:L-kin}-\eqref{eq:zeta}].
The canonical graviton field $\,h_{\mu\nu}^{}\,$ is shifted
by a linear term of $\,\phi\,$ from the original field $\,\hat{h}_{\mu\nu}^{}\,$ \cite{XRH},\,
but this does not affect the couplings of $\,h_{\mu\nu}^{}\,$ with dark matter field
$\,\chi_s^{}\,$ and other SM fields.}
\begin{eqnarray}
\label{eq:chischishmn}
\frac{\,\sqrt{2}\,}{\Mp}\left[\,\xis\!\(p^{\mu}p^{\nu}-p^2\eta^{\mu\nu}\)
 +\(p_1^{(\mu}p_2^{\nu)}-\frac{1}{2}p_1^{}\!\cdot p_2^{}\eta^{\mu\nu}\)\right]\!,
\end{eqnarray}
where the first term comes from nonminimal coupling and is proportional to $\,\xis\,$.\,
The SM particles couple to gravity minimally through their energy-momentum tensor.
The cubic couplings of a pair of SM particles with $\,h_{\mu\nu}^{}\,$
are suppressed by $\,\Mp^{-1}$\,.\,
In the parameter region of $\,|\xis|\gg 1$\,,\, the interactions between GDM and
SM particles induced by graviton-exchange are largely enhanced.
Furthermore, since Higgs field mixes with graviton via kinetic term,
$\,\X$ can communicate with SM particles via Higgs-exchange.
Such contributions are proportional to $\,\xis\xih\,$,\,
which will be much more enhanced when both $\,|\xis|,|\xih|\gg 1\,$.\,
The explicit momentum structures of these interactions are
determined by the complicated tensor structure of graviton propagator and the related vertices.
We note that the analysis will be much simplified by transformation into Einstein frame.
In the following, we will explicitly derive the new set of
effective Feynman vertices involving the GDM
interactions with the SM particles in Einstein frame.

\vspace*{3mm}
\subsection{\hspace*{-3.5mm} Minimal GDM in Einstein Frame}
\label{sec:2.2}
\vspace*{2mm}

The Einstein frame is defined by the conventional metric that satisfies Einstein equation.
This is achieved by eliminating non-minimal coupling terms via the Weyl transformation.
For notational convenience, we will suppress the superscript ``$(E)$" for geometric quantities
in Einstein frame. The Weyl transformation is defined as,
$\,g_{\mu\nu}^{} = \Omega^2g^{(J)}_{\mu\nu}$,\, and the factor $\,\Omega^2\,$ is given by
\begin{eqnarray}
\label{Omega}
\Omega^2 ~=~
\frac{\,M^2+2\xih H^\dag H+\xis\chi_s^2\,}{\Mp^2}
~=~ 1+\frac{\,\xih (2\vew\hat\phi+\hat\phi^2+|\pi|^2)+\xi_s\chi_s^2\,}{\Mp^2}\, ,
\hspace*{5mm}
\end{eqnarray}
where $\,|\pi|^2 = 2\pi^+\pi^-+(\pi^0)^2$\,.\,
Accordingly, the Weyl transformation of Ricci scalar takes the following form,
\begin{eqnarray}
\mathcal{R}^{(J)}\,=~
\Omega^2\Big[\mathcal{R}-6g^{\mu\nu}\nabla_\mu\nabla_\nu\log\Omega+6g^{\mu\nu}
\big(\nabla_\mu\log\Omega\big)\big(\nabla_\nu\log\Omega\big)\Big]\, .
\end{eqnarray}
Substituting this into (\ref{SRNMCJF}), we derive the Einstein frame action for bosonic sector,
\begin{eqnarray}
\label{SRNMCEFb}
S_{\text{E}}^{\text{b}}
&=& \!\int\!\! \textrm{d}^4x\,\sqrt{-g}\left\{\frac{1}{2}\Mp^2 \mathcal{R}
-\frac{1}{4}F_{\mu\nu i}^aF^{\mu\nu a}_i
+\frac{3}{\Mp^2\Omega^4}\left[\partial_\mu^{}\!\left(\xih H^\dag H
+\frac{1}{2}\xis\XX\right)\right]^2
\right.
\hspace*{10mm}
\nonumber\\
    && \hspace*{20mm} \left.
    +\frac{1}{\Omega^2}(D_\mu H)^\dag(D^\mu H)+\frac{1}{2\Omega^2}\partial_\mu^{}\X\partial^\mu_{}\X
    -\frac{1}{\Omega^4}V(H,\X )
\right\} .
\end{eqnarray}
For nonzero $\,\xi_h^{}\,$,\, the higher dimensional operator in the first line of (\ref{SRNMCEFb})
yields additional contribution to the Higgs kinetic term.
Together with the original one, we have the following kinetic term for the Higgs and Goldstone boson fields,
\begin{eqnarray}
\label{eq:L-kin}
\mathcal{L}_{\textrm{kin}}^{} =
\frac{1}{2}\left(1+\frac{\,6\xi_h^2\vew^2\,}{\Mp^2}\right)\!
(\partial_\mu \hat\phi)^2+\partial_\mu \pi^+\partial^\mu \pi^-+\frac{1}{2}(\partial_\mu \pi^0)^2\, .
\end{eqnarray}
Hence, we can normalize the kinematic term of Higgs boson by a field redefinition,
$\, \hat\phi = \zeta\phi\,,$\,
with the rescaling factor,
\beqa
\label{eq:zeta}
\zeta \,=\, \(1+\frac{\,6\xi_h^2\vew^2\,}{\Mp^2}\)^{\!\!-\frac{1}{2}}_{}\,.
\eeqa
Then, the canonical field \,$\phi$\, is identified as the 125\,GeV Higgs boson,
which was recently discovered at the LHC \cite{LHC-h125,LHC2014}.
We note that this rescaling only applies to the Higgs field $\,\hat\phi\,$,\, but
does not affect its constant vacuum expectation value $\,\vew^{}\,$.\,
The same operator in \eqref{SRNMCEFb} also induces self-interactions for scalars.

\vspace*{1mm}

For the fermionic sector, we write down the pure kinetic term and mass-term
for a generic Dirac spinor $\,f\,$ (quark or lepton) in Jordan frame,
\beqa
S_{\text{F}}^{}
\,= \int\!\!\di^4x\,\det(e_\nu^q) \left[ \bar{f} \ga^p e^\mu_p
\Big(\ii\pd_\mu-\fr{1}{2}\omega_\mu{}^{mn}\si_{mn}\Big) f
- m_f^{}\bar{f} f
\right] \!,
\eeqa
where $\,e_\nu^q\,$ and $\,\omega_\mu^{}{}^{mn}\,$ denote the vierbein and spin-connection,
and $\,\si_{mn}=\fr{\ii}{2}[\ga_m,\ga_n]$\,.\,
Setting the flat background in Einstein frame, we deduce the metric in Jordan frame,  $\,g_{\mu\nu}^{(J)}=\Omega^{-2}\eta_{\mu\nu}^{}\,$.\,
Thus, we can express the vierbein and spin-connection in Jordan frame as functions of $\,\Omega\,$,
\begin{eqnarray}
e_{\mu}^{m} ~=~ \Omega^{-1}\delta_\mu^m \,,
\qquad
\omega_{\mu}^{}{}^{mn} \,=\, -\Omega^{-1}
(\delta_\mu^m\partial^n\Omega-\delta_\mu^n\partial^m\Omega)\, .
\end{eqnarray}
With these, we can explicitly write down the kinetic term and mass-term for the SM fermions
(quarks or leptons) in the Einstein frame \cite{XRH},
\begin{eqnarray}
\label{SRNMCEFf}
S_{\text{E,f}} ~=\int \!\!\textrm{d}^4x\,\sqrt{-g}
\left[\frac{1}{\Omega^3}\!\left(\!\bar{f}\,\ii\slashed\partial\, f
 +\frac{3}{\Omega}\bar{f}\,(\ii\slashed\partial\Omega) f \!\right)
 -\frac{m_f^{}}{\Omega^4}\bar{f}f \right]\! .
\end{eqnarray}

In the following, we summarize the vertices relevant for DM annihilation processes,
by expanding $\,\Omega\,$ at the leading order of $\,1/\Mp^2\,$.

\vspace*{4mm}
\noindent
\underline{$\bullet$  GDM Interactions with Higgs and Goldstone Bosons:~}
\vspace*{3mm}

  The couplings of $\,\X\,$ to Higgs and Goldstone bosons depend on both $\,\xih\,$ and $\,\xis\,$.\,
  From Eq.\,(\ref{SRNMCEFb}), we summarize these interaction terms as follows,
  \begin{eqnarray}
  \label{Lssint}
  \mathcal{L}_{\textrm{int}}^{ss}
  &=&
  \frac{3}{\,4\Mp^2\,}
  \left\{\xi_h^2\Big[\partial_\mu^{}\!\(2\vew\zeta\phi+\zeta^2\phi^2\!+\! |\pi|^2\)\!\Big]^2
  + 4\xih \xis \X\partial^\mu \X\partial_\mu^{}
  \!\(2\vew\zeta\phi+\zeta^2\phi^2 \!+\! |\pi|^2\) \right.
  \nonumber\\
  &&
  +4\xi_s^2\XX (\partial_\mu \X)^2\Big\}-\frac{1}{2}\left[\frac{\xih}{\Mp^2}
  \left(2\vew\zeta\phi+\zeta^2\phi^2 \!+\! |\pi|^2\right)+\frac{\xis}{\Mp^2}\XX\,\right]\!\times
  \nonumber\\
  &&
  \Big[\zeta^2(\partial_\mu \phi)^2 \!+\! |\partial_\mu \pi|^2+(\partial_\mu\X)^2\,\Big]\, ,
  \end{eqnarray}
  where $\,\phi\,$ is the canonical Higgs field and
  $\,\zeta=(1+6\xi_h^2\vew^2/\Mp^2)^{-1/2}$\, is the rescaling factor given by Eq.\,\eqref{eq:zeta}.
  We also have,
  $|\partial_\mu^{} \pi|^2 = 2\partial_\mu^{} \pi^+\partial^\mu \pi^- \!+ (\partial_\mu^{}\pi^0)^2$.\,
  Note that the interactions in the first brackets $\{\cdots\}$ are induced
  by higher dimensional operator in the first line of Eq.\,(\ref{SRNMCEFb}),
  which includes quadratic terms of $\,\xih\,$ and $\,\xis\,$.\,
  The other terms arise from expanding $\,1/\Omega^2\,$ for scalar kinetic terms,
  which only depend on $\,(\xih,\, \xis)$\, linearly.
  Hence, for $\,(\xih,\,\xis) \gg 1$\,,\, the quadratic terms of $\,(\xih,\, \xis)$\,
  will make dominant contributions.

\vspace*{1mm}

  The only triple coupling relevant to the following analysis comes from
  the vertex $\,\X\!-\!\X\!-\phi$\,.\,
  It induces Higgs invisible decay when $\,\MX < \frac{1}{2}m_{\phi}^{}$\,,\,
  and also generates interactions between the GDM and SM particles by exchanging the Higgs boson.
  We derive the corresponding Feynman vertex at the leading order,
  \begin{eqnarray}
  \label{eq:FVsXX}
  \X (p_1^{})\!-\!\X (p_2^{})\!-\!\phi(q)\!:~~~
    \ii \frac{\,2\xih \zeta \vew\,}{\Mp^2}(p_1^{}\cdot p_2^{})
  + \ii \frac{\,6\xih \xis \zeta \vew\,}{\Mp^2}q^2\, ,
  \hspace*{10mm}
  \end{eqnarray}
  where all momenta flow inwards.
  Then, we deduce the quartic couplings between the GDM $\,\X\,$ and Higgs/Goldstone bosons
  at the leading order,
  \begin{eqnarray}
  \label{eq:FVssXX}
  &&
  \X (p_1^{})\!-\!\X (p_2^{})\!-\!\pi^{+,0}(p_3^{})\!-\!\pi^{-,0}(p_4^{})\!:~~~
  \ii\frac{2}{\Mp^2}\Big[ 3\xih \xis q^2+\xih (p_1^{}\cdot p_2^{})+\xis (p_3^{}\cdot p_4^{})\Big] ,
  \hspace*{10mm}
  \nonumber\\
  &&
  \X (p_1^{})\!-\!\X (p_2^{})\!-\!\phi(p_3^{})\!-\!\phi(p_4^{})\!:~~~
  \ii\frac{2\zeta^2}{\Mp^2}\Big[ 3\xih\xis q^2+\xih (p_1^{}\cdot p_2^{})+\xis (p_3^{}\cdot p_4^{})\Big] ,
  \hspace*{10mm}
  \end{eqnarray}
  where $\,q=p_1^{}+p_2^{}\,$.\,
  They also contribute to the dark matter annihilations in early universe and today.
  To obtain leading order contributions form Higgs exchange to these vertices,
  we need the following triple couplings,
  \beq
  \ba{ll}
  \phi-\phi-\phi\!: &~~~\dis -\ii\frac{\,3m_\phi^2\,}{\vew}\zeta^3 \,,
  \\[4mm]
  \pi^{+(0)}\!-\pi^{-(0)}\!-\phi\!: &~~~\dis -\ii\frac{\,m_\phi^2\,}{\vew}\zeta \,.
  \ea
  \eeq
  Then, we deduce the quartic coupling for the vertex
  $\,\X (p_1^{})\!-\!\X (p_2^{})\!-\!\pi^{+,0}(p_3^{})\!-\!\pi^{-,0}(p_4^{})$\,
  at the leading order,
  \begin{eqnarray}
  \ii\frac{1}{\Mp^2}\!\left[6\xih \xis q^2+2\xih(p_1^{}\cdot p_2^{})+2\xis(p_3^{}\cdot p_4^{})
   +\xih m_\phi^2\zeta^2\frac{\,(6\xis \!+\!1)q^2\!-\!2M_\chi^2\,}{\,q^2\!-m_\phi^2\!
   +\ii m_\phi^{}\Gamma_\phi^{}\,}\right]\! ,
  \hspace*{9mm}
  \end{eqnarray}
  where $\,\Gamma_{\phi}^{}\,$ stands for the Higgs boson width.
  For the quartic coupling with Higgs bosons, \,$t(u)$-channel exchange of $\,\X\,$
  also contribute, and the vertex
  $\,\X (p_1^{})\!-\!\X (p_2^{})\!-\!\phi(p_3^{})\!-\!\phi(p_4^{})$\, becomes
  \begin{eqnarray}
  \label{eq:FVppXX}
  && \ii\frac{\zeta^2}{\Mp^2}\!\left[6\xih \xis q^2+2\xih(p_1^{}\cdot p_2^{})+2\xis(p_3^{}\cdot p_4^{})
  +3\xih m_\phi^2\zeta^2
   \frac{\,(6\xis\!+\!1)q^2\!-\!2M_\chi^2\,}
        {\,q^2\!-\!m_\phi^2\!+\!\ii m_\phi\Gamma_\phi^{}\,}  \right.
  \nonumber\\[2mm]
  && \left.
  -\frac{\xi_h^2 \vew^2}{\Mp^2}\left(\frac{\((6\xis+1)m_\phi^2-M_\chi^2-t\)^2}{t-m_\phi^2+\ii  m_\phi\Gamma_\phi}+\frac{\((6\xis+1)m_\phi^2-M_\chi^2-u\)^2}{u-m_\phi^2+\ii m_\phi\Gamma_\phi}\right)
  \right]\!.
  \hspace*{9mm}
  \end{eqnarray}
  where $\,t=(p_1-p_3)^2\,$ and $\,u=(p_1-p_4)^2$.\,
  The quartic couplings for the $\,4\phi$\, and $\,4\X$ vertices
  as well as for the Higgs-Goldstone interactions receive quite similar contributions.
  They will be included in our coupled channels analysis of perturbative unitarity.

\vspace*{4mm}
\noindent
\underline{$\bullet$  GDM Interactions with Weak Gauge Bosons:~}
\vspace*{3mm}

  Under Weyl transformation the gauge boson kinetic terms remain intact as in Eq.\,(\ref{SRNMCEFb}).
  We note that the tree-level interactions between the GDM and massive gauge bosons
  arise from the gauge boson mass-term.
  For weak gauge bosons, this is associated with the Higgs kinetic term in Eq.\,(\ref{SRNMCEFb}).
  Thus, we derive the interaction term,
  $\, -\frac{\,\xis m_V^2\,}{2\Mp^2}\delta_V^{} V^\mu V_\mu\chi_s^2$\,,\,
  with the notation $\,V\in (W,\,Z)$\, and coefficients
  $\,(\delta_W^{},\, \delta_Z^{})=(2,\,1)$\,.\,
  Hence, we infer the Feynman vertex of gravity-induced contact interaction for
  the GDM and weak bosons,
  \begin{eqnarray}
  \label{eq:FVVVXXc}
  \X\!-\!\X\!-\!V_\mu^{}-\!V_\nu^{}\!:~~
  -\ii\frac{\,2\xis m_V^2\,}{\Mp^2} g^{\mu\nu} \, .
  \hspace*{10mm}
  \end{eqnarray}
  Besides, the GDM can interact with weak bosons by exchanging the Higgs boson.
  With the gravity-induced $\,\X\!-\X\!-\phi$\, vertex in Eq.\,(\ref{eq:FVsXX})
  and the $\,V_\mu^{}\!-\!V_\nu^{}\!-\!\phi$\, vertex from the SM,
  we derive the following contribution via Higgs-exchange at the leading order,
  \begin{eqnarray}
  \label{eq:FVVVXXh}
  \X (p_1^{})\!-\!\X (p_2^{})\!-\!V_\mu^{}\!-\!V_\nu^{}\!:~~~
  -\ii \xih \zeta^2
  \frac{\,(6\xis \!+\!1)q^2\!-\!2M_\chi^2\,}
       {q^2-m_\phi^2+\ii m_\phi\Gamma_\phi}\frac{\,2m_V^2\,}{\Mp^2}g^{\mu\nu}\, ,
  \end{eqnarray}
  where $\,q=p_1^{}+p_2^{}\,$.\,
  Then, we deduce an effective (nonlocal) vertex for
  $\,\X (p_1^{})\!-\!\X (p_2^{})\!-\!V_\mu^{}\!-\!V_\nu^{}\,$ as follows,
  \begin{eqnarray}
  \label{eq:FVVVXX}
  -\ii g^{\mu\nu}\frac{\,2m_V^2\,}{\Mp^2}\left[\xis + \xih \zeta^2
  \frac{\,(6\xis \!+1)q^2\!-2M_\chi^2\,}{\,q^2\!-m_\phi^2+\ii m_\phi^{}\Gamma_\phi^{}\,}\right] .
  \end{eqnarray}
  For $\,|\xih|, |\xis| \gg 1\,$,\,
  it is the quadratic term of $\,\xih\xis\,$ in Eq.\,(\ref{eq:FVVVXXh})
  that will make dominant contribution.
  For the scattering process $\,V_L^{}V_L^{}\to\X\X\,$,\,
  the non-renormalizable gravity-induced interactions will contribute a net $E^2$-dependence in the amplitude,
  and cause perturbative unitarity violation at high energies.
  Furthermore, this vertex will lead to the GDM pair-productions via weak boson scattering
  $\,VV\to\X\X\,$ at the LHC and future high energy $pp$ colliders.

\vspace*{4mm}
\noindent
\underline{$\bullet$  GDM Interactions with Fermions:~}
\vspace*{3mm}

  According to Eq.\,(\ref{SRNMCEFf}), the dark matter can interact with fermions
  via their kinetic terms or mass-terms. Intuitively, the kinetic terms in the parentheses seem
  to induce momentum-dependent higher dimensional operators with
  $\,\Omega^{-4} \simeq 1 -2\XX /M_*^2$.\,
  But, for on-shell fermions, the contributions from kinetic terms share the same structure
  as that from mass-terms. The total contribution to the contact interaction is
  $\,\frac{\,\xis m_f^{}\,}{\,\Mp^2\,}\bar{f}f\chi_s^2$\,.\,
  In addition, the Higgs-exchange induces a nonlocal contribution to the same vertex at the leading order.
  Thus, we explicitly derive the Feynman vertex $\,\X (p_1^{})\!-\!\X (p_2^{})\!-\!\bar{f}\!-\!f$\,
  with effective coupling,
  \begin{eqnarray}
  \label{eq:FVqqXX}
  \ii \frac{m_f^{}}{\,\Mp^2\,}\!
  \left[\,\xis + \xih \zeta^2
  \frac{\,(6\xis \!+\!1)q^2\!-\!2M_\chi^2\,}{\,q^2\!-\!m_\phi^2\!+\!\ii m_\phi^{}\Gamma_\phi^{}\,}\,\right]\!.
  \end{eqnarray}
  We note that the terms in the brackets of \eqref{eq:FVqqXX} and \eqref{eq:FVVVXX} take the same form.
  At high energies, the scattering amplitude of $\,\X\X\to\bar{f}f\,$ contains non-canceled leading $E^1$ terms,
  which will eventually violate perturbative unitarity as the energy $\,E$\, increases \cite{AC,Unitarity}.

\vspace*{4mm}
\noindent
\underline{$\bullet$ GDM Interactions with Massless Gauge Bosons:~}
\vspace*{3mm}

  \begin{figure}[h]
  \centering%
  \includegraphics[width=12.5cm]{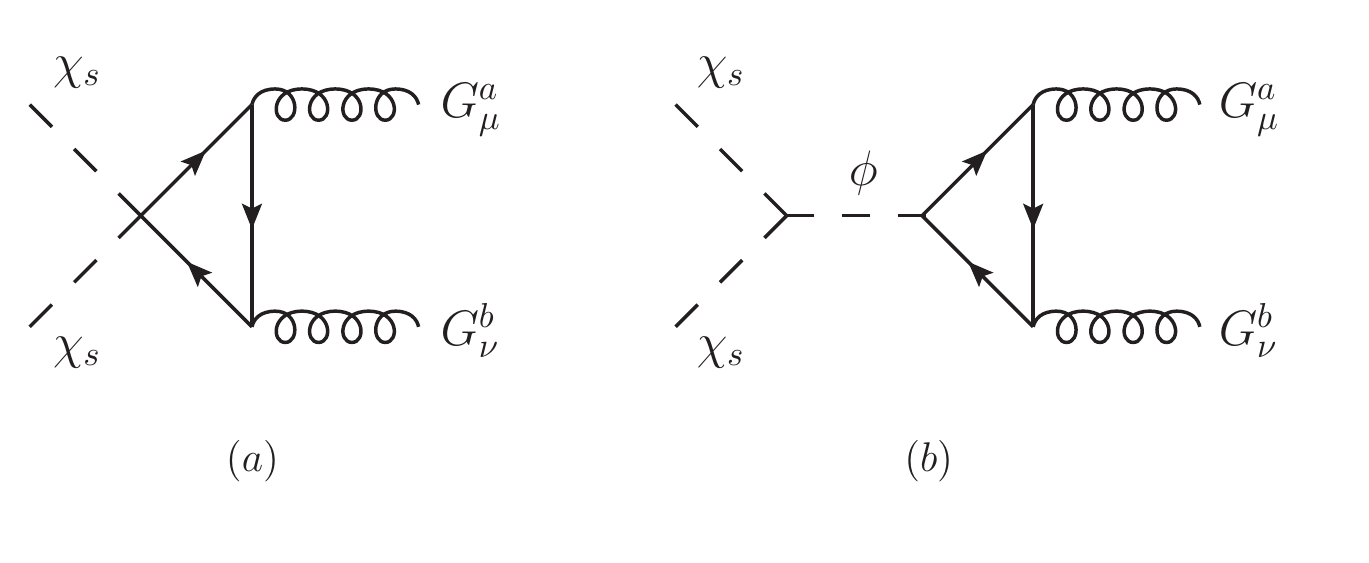}
  \vspace*{-6mm}
  \caption{One-loop diagram for the dimension-6 effective operator $\,\chi_s^2 G^a_{\mu\nu}G^{a\mu\nu}$\,.}
  \label{fig:1}
  \end{figure}

   As shown in Eq.\,(\ref{SRNMCEFb}), the gauge boson kinetic terms remain intact under Weyl transformation.
   So there is no contact interaction of the GDM with massless gauge bosons (gluons or photons)
   at the leading order. Nevertheless, there are loop-induced higher dimensional operators.
   For instance, the dimension-6 operator $\,\chi_s^2 G^a_{\mu\nu}G^{a\mu\nu}$\,
   can be generated by the top quark triangle-loop in Fig.\,\ref{fig:1},
   where the diagram (a) involves leading order $\,\X \X \bar{f}f$\, contact interaction,
   and the diagram (b) includes Higgs-exchange with Higgs effective coupling to gluons.
   It will initiate gluon-fusion production of $\,\X\X\,$ at the LHC and the future high energy hadron colliders.
   In parallel, the dimension-6 operators $\,\chi_s^2 A_{\mu\nu}^{}A^{\mu\nu}$\, and
   $\,\chi_s^2 A_{\mu\nu}^{}Z^{\mu\nu}_{}$\, can be generated from both $W^\pm$ loop and fermion loop,
   and are relevant to the indirect detections of dark matter.
   Inspecting (\ref{eq:FVVVXX}) and (\ref{eq:FVqqXX}), we note the similarity between the contact vertex
   for $\,\chi_s^2VV\,$ ($\,\chi_s^2\bar{f}f\,$) and the corresponding
   $\,\phi VV\,$ (\,$\phi\bar{f}f$\,) vertex. With the same structure,
   couplings of the former can be reproduced from the latter by
   the substitution $\,\vew\to -\Mp^2/\xis\,$.\,
   Hence, we can directly infer the form of these one-loop generated vertices
   from the conventional results for the SM Higgs boson \cite{SMHRev} as follows,
  \begin{eqnarray}
  \label{eq:FVAAXX}
  &&
  \X \!-\! \X \!-\! G^a_\mu (p_3^{}) \!-\! G^b_\nu (p_4^{})\!:
  \nonumber\\
  &&
  \qquad\qquad
  \ii\,\mathcal{C}_g\frac{2\alpha_s}{3\pi\Mp^2}\Big((p_3^{}\cdot p_4^{})g^{\mu\nu}-p_3^\nu p_4^\mu\Big)
  \left[\xis + \xih \zeta^2
  \frac{\,(6\xis\!+\!1)q^2\!-2M_\chi^2\,}
  {\,q^2\!-m_\phi^2+\ii m_\phi^{}\Gamma_\phi^{}\,}\right]\! ,
  \qquad\nonumber
  \\[2mm]
  &&
  \X \!-\! \X \!-\! A_\mu^{} (p_3^{})\!-\! A_\nu^{} (p_4^{})\!:
  \nonumber\\
  &&
  \qquad\qquad
  \ii\,\mathcal{C}_\gamma\frac{8\alpha}{\pi\Mp^2}
  \Big((p_3^{}\cdot p_4^{})g^{\mu\nu}-p_3^\nu p_4^\mu\Big)
  \left[\xis + \xih \zeta^2\frac{\,(6\xis \!+\!1)q^2-2M_\chi^2\,}{\,q^2\!-m_\phi^2
   +\ii m_\phi^{}\Gamma_\phi^{}\,}\right]\!,
  \qquad
  \\[2mm]
  &&
  \X \!-\! \X \!-\! A_\mu (p_3^{}) \!-\! Z_\nu^{} (p_4^{})\!:
  \nonumber\\
  &&
  \qquad\qquad
  \ii\,\mathcal{C}_{\gamma z}\frac{4\alpha}{\pi\Mp^2}
  \Big((p_3^{}\cdot p_4^{})g^{\mu\nu}-p_3^\nu p_4^\mu\Big)
  \left[\!\xis + \xih \zeta^2\frac{\,(6\xis \!+\!1)q^2\!-2M_\chi^2\,}
                                  {\,q^2\!-m_\phi^2\!+\ii m_\phi^{}\Gamma_\phi^{}\,}\right]\!,
  \qquad
  \nonumber
  \end{eqnarray}
  where the form factors \,$(\mathcal{C}_g,\, \mathcal{C}_\gamma,\, \mathcal{C}_{\gamma z})$\,
  are energy-dependent,
  \begin{eqnarray}
  \label{eq:loopfactors}
  \mathcal{C}_g &\,=\,&
  A_F^{}(\tau_t^{})+A_F^{}(\tau_b^{})+A_F^{}(\tau_c^{}) \,,
  \nonumber\\[1mm]
  \mathcal{C}_\gamma
  &=& -A_V^{}(\tau_W^{})+\fr{1}{18}A_F^{}(\tau_b^{})+\fr{2}{9}
      [A_F^{}(\tau_t^{})+A_F^{}(\tau_c^{})]+\fr{1}{6}A_F^{}(\tau_\tau^{}) \,,
  \\[1mm]
  \mathcal{C}_{\gamma z} &=& B_V^{}(\tau_W^{}, \eta_W^{})
  +B_F^{}(\tau_t^{}, \eta_t^{})+B_F^{}(\tau_b^{}, \eta_b^{})
  +B_F^{}(\tau_c^{}, \eta_c^{})+B_F^{}(\tau_\tau^{}, \eta_\tau^{}) \,,
  \hspace*{10mm}
  \nonumber
  \end{eqnarray}
  where $\,\tau_j^{}=q^2/4m_j^2$\, with $\,q=p_3^{}+p_4^{}\,$,\, and
  $\,\eta_j^{} = m_Z^2/4m_j^2$\,.\,
  The explicit expressions of $\,A_{V,F}^{}(\tau)\,$ and $\,B_{V,F}^{}(\tau, \eta )\,$
  are given in Appendix\,\ref{app:loopfactor}.
  For analysis of non-relativistic dark matter annihilations in Sec.\,\ref{sec:3},
  we will have $\,q^2\approx 4M^2_\chi\,$.

\vspace*{4mm}
\subsection{\hspace*{-3.5mm} Perturbative Unitarity}
\label{sec:2.3}
\vspace*{2mm}

In this subsection, we derive perturbative unitarity bound from high energy scattering processes involving the GDM,
as induced by the non-renormalizable gravitational interactions in Sec,\,\ref{sec:2.2}.
For gauge bosons and the Goldstone bosons, the leading order amplitudes in the high energy limit
are given by $\,\OO(E^2)\,$ terms.
Thus, we derive the following amplitudes,
\begin{eqnarray}
\label{eq:XX-VV/hh}
\mathcal{T}[\X\X\!\to V_L^{a}V_L^{a}]
&~\simeq\,& -\mathcal{T}[\X\X \!\to \pi^{a}\pi^{a}\,]
\nonumber\\[2mm]
&~\simeq\,& -\frac{E^2}{\,\Mp^2\,}\(6\xih \xis\!+ \xih\! +\xis\) + \mathcal{O}(E^0)\, ,
\end{eqnarray}
where $\,E=\sqrt{s}\,$ is the center of mass energy of the scattering.
Here, we keep the leading order contributions at $\,\OO(1/\Mp^2)$\,.\,
We compute the scattering amplitudes for both the longitudinal gauge boson final state $\,V_L^aV_L^a\,$
and the corresponding Goldstone boson final state $\,\pi^a\pi^a\,$.\,
This verifies the equivalence theorem\,\cite{ET} at high energies and serves as nontrivial consistency checks
of our analysis.  For the Higgs final state, we find that the leading amplitude is given by
$\,\mathcal{T}[\X\X \!\to \phi\phi\,] \simeq \mathcal{T}[\X\X \!\to \pi^{a}\pi^{a}\,]\,$,\,
in the high energy regime, which arises from the contact interaction (\ref{eq:FVssXX})
at $\,\OO(1/\Mp^2)$\,.

\vspace*{1mm}

To derive the optimal perturbative unitarity constraint, we further perform the coupled channel analysis
for the normalized two-body scalar states,
$|\pi^+\pi^-\rangle$, $\fr{1}{\sqrt{2}\,}|\pi^0\pi^0\rangle$,
$\frac{1}{\sqrt{2}\,}|\phi\phi\rangle$,
$|\pi^0\phi\rangle$, $\frac{1}{\sqrt{2}\,}|\X\X\rangle$, $|\pi^0\X\rangle$ and $|\phi\X\rangle$.\,
The partial wave amplitude is given by
\begin{eqnarray}
a_\ell^{}(E) \,=\,
\frac{1}{\,32\pi\,}\!\int_{-1}^1\!\! d\cos\theta\, P_\ell^{}(\cos\theta)\mathcal{T}(E,\theta)\,.
\end{eqnarray}
We inspect the leading contributions at $\,\OO(E^2/\Mp^2)$\,.\,
The leading amplitudes without involving $\,\X\,$ were derived before in Ref.\,\cite{XRH}.
Combined these with the amplitudes of (\ref{eq:XX-VV/hh}) and related results,
we deduce the full $s$-wave amplitude in matrix form,
\begin{eqnarray}
\hat{\bf a}_0^{} &=&
\left(\!\begin{array}{ccc}
\AA_{11}^{} & \AA_{12}^T & {\bf 0}
\\[1.5mm]
\AA_{12}^{} & {0} & {\bf 0}
\\[1.5mm]
{\bf 0} & {\bf 0} & \AA_{33}^{}
\end{array}
\right)\! .
\label{eq:a0-hat}
\end{eqnarray}
The submatrices in \eqref{eq:a0-hat} take the following form,
\beqs
\label{eq:A11-12-33}
\begin{eqnarray}
\label{eq:A11}
\AA_{11}^{} & \simeq &  \frac{3\xi_h^2E^2}{\,16\pi\Mp^2\,}\!\left(\!
\begin{array}{cccr}
1 & \sqrt{2}~ & \sqrt{2}~ & {0}
\\[1mm]
\sqrt{2}~ & {0} & 1 & {0}
\\[1mm]
\sqrt{2}~ & 1 & {0} & {0}
\\[1mm]
{0} & {0} & {0} & -1
\\[1mm]
\end{array}
\,\right) \!,
\\
\AA_{12}^{} & \simeq & \frac{\,3\xih\xis\eta^{\frac{1}{2}}_{12}E^2\,}{16\pi\Mp^2}\!
\left(\!\sqrt{2},\, 1,\, 1,\, 0 \right) \!,
\quad
\AA_{33}^{} \,\simeq\, \frac{\,3\xih\xis\eta^{}_{33} E^2\,}{8\pi\Mp^2}\,
\textrm{diag}(1,\,1) \,,
\hspace*{10mm}
\label{eq:A12-33}
\end{eqnarray}
\eeqs
where we only keep the leading $E^2$-terms under the limit
$\,|\xih|,|\xis| \gg 1$\,.\,
In the above formulas \eqref{eq:A12-33},
$\,\eta_{12}^{}=(1-4M_\chi^2/E^2)^{\frac{1}{2}}_{}$\, and
$\,\eta_{33}^{}=(1-M_\chi^2/E^2)$.\,
Here, for the convenience of applying the unitarity conditions below, we have included
proper kinematical phase factor of each scattering channel (such as $\,\eta_{12}^{}$\,
and $\,\eta_{33}^{}$\,), which were generally defined in Appendix\,B
of the first paper in \cite{Unitarity}.
In the present unitarity analysis, it suffices to keep only the mass $\,\MX\,$
(which could reach TeV scale) and ignore other small masses of weak bosons and Higgs boson
in comparison with the large scattering energy $\,E\,$.\,
Thus, in Eq.\,\eqref{eq:A11-12-33}, only the scattering channels involving external $\X$ state
have nontrivial phase factor $\,\eta_{ij}^{}\neq 1\,$.\,
After diagonalization, we deduce the eigenvalue amplitudes,
\begin{eqnarray}
\hat{\bf a}_{0,\textrm{diag}}^{}
\,\simeq\, \frac{3\xih E^2}{\,16\pi\Mp^2\,}
\textrm{diag}( x_1^{},\,  x_2^{},\, -\xih,\, -\xih ,\, -\xih,\,
 2\eta_{33}^{}\xis,\, 2\eta_{33}^{}\xis )\, ,
\end{eqnarray}
where
$\,x_{1,2}^{}=\frac{1}{2}\!
 \left(3|\xih |\pm\!\sqrt{9\xi_h^2\!+\!16\eta_{12}^{}\xi_s^2\,}\,\right)$.\,
The $s$-wave amplitude should obey the unitarity condition
$\,|\hat{a}_0^{}|<1\,$ (or, $\,|\text{Re}\hat{a}_0^{}|<1/2\,$) \cite{Unitarity}.
Imposing condition $\,|\hat{a}_0^{}|<1\,$
on the maximal eigenvalue, we derive the unitarity bound $\,\ucut = E_{\max}^{}\,$,
\begin{eqnarray}
\label{eq:PUBE2}
E ~<~ \ucut =\, \min\!\(\!\dis
\frac{\,\sqrt{32\pi}\Mp^{}\,}
{\,\left[3|\xih | \!\!
 \(3|\xih |\!+\!\sqrt{9\xi_h^2\!+\!16\bar{\eta}_{12}^{}\xi_s^2}\,\)\!\right]^{1/2}\,},\,
\frac{\sqrt{8\pi}\Mp^{}}{\,\sqrt{3\bar{\eta}_{33}^{}\xih\xis\,}\,}
\!\)\!, \hspace*{5mm}
\end{eqnarray}
where
$\,\bar{\eta}_{12}^{}=(1-4M_\chi^2/\ucutt )^{\frac{1}{2}}_{}$\, and
$\,\bar{\eta}_{33}^{}=(1-M_\chi^2/\ucutt )$.
Defining the coupling ratio, $\,r\equiv |\xis /\xih |\,$,\,
we can express \eqref{eq:PUBE2} as an upper bound on $\,\sqrt{|\xih\xis |\,}\,$
for each given energy $E$,
\beqa
\label{eq:xih-xis-UB}
\sqrt{|\xih\xis |} ~<\,
\,\min\!\(\!\frac{\sqrt{8\pi/3\,}}
  {\,\left[\!\sqrt{\eta_{12}^{}\!+\!\(\!\frac{3}{4r}\!\)^2\,}\!+\!\frac{3}{4r}\right]^{\!{1}/{2}}\,},\,
  \sqrt{\frac{8\pi}{3\eta_{33}^{}}}\!\)
  \!\!\frac{\,\Mp^{}\,}{E} \,.
\eeqa
Here, the strongest limit corresponds to $\,E=E_{\max}^{}=\ucut$,\,
which serves as an ultraviolet (UV) cutoff of this effective theory.
In our present study, we will set up the parameter space
$\,|\xis | > |\xih | \gg 1\,$,\,
where typically we take the coupling ratio $\,r = 5-30\,$.\,
It is clear that for the range of $\,r = 5-30\,$,\,
we have $\,r^{-2}\ll 1\,$ and thus the bound \eqref{eq:xih-xis-UB} is not so sensitive
to the ratio $\,r\,$.\,

\vspace*{1mm}

Besides, since $\,\xi_s^2>0$\, and $\,\xi_h^2>0$\, in \eqref{eq:PUBE2},
we can always derive an upper bound on $\,\xih\,$ alone
(for each given energy scale $E$\,),\,
\beqa
\label{eq:xih-UB}
|\xih | ~&<&~ \frac{\sqrt{16\pi}\,}{3}\frac{\,\Mp^{}\,}{E} \,.
\eeqa

We further note that the scattering amplitude of $\,\X\X \to \X\X$\,
vanishes at $\,\mathcal{O}(E^2)\,$  due to the crossing symmetry,
we may further consider its subleading terms at $\,\mathcal{O}(E^0)\,$,\,
which is still enhanced by $\,\xi_s^2\,$.\,
From the Lagrangian (\ref{Lssint}), we derive the following amplitude,
\begin{eqnarray}
\mathcal{T}\!\left[\fr{1}{\sqrt{2}\,}\X \X \!\to\! \fr{1}{\sqrt{2}\,}\X \X\right]
\,=\, \frac{\,2\xis (6\xis \!-\!1)M_\chi^2\,}{\Mp^2}\, ,
\end{eqnarray}
which leads to the $s$-wave amplitude,
\beqa
a_0^{}\!\left[\fr{1}{\sqrt{2}\,}\X \X \!\to\! \fr{1}{\sqrt{2}\,}\X \X\right]\,\simeq\,
\frac{\,3\xi_s^2}{\,4\pi\,} \!\(\!\frac{\MX}{\Mp^{}}\!\)^{\!\!2}\, ,
\eeqa
for $\,\xis \gg 1\,$.\,
Thus, imposing the unitarity condition $\,|\bar{\eta}_{12}^{}a_0^{}|<1\,$ \cite{Unitarity},
we deduce the following bound
for $\,\xis \gg 1\,$,\,
\begin{eqnarray}
\label{eq:PUBE0}
|\xis | \,<\, \sqrt\frac{4\pi}{\,3\bar\eta_{12}^{}\,}\frac{\Mp^{}}{M_\chi^{}}\, ,
\hspace*{8mm}
\text{or,}
\hspace*{8mm}
M_\chi^{} \,<\, \sqrt\frac{4\pi}{\,3\bar\eta_{12}^{}\,}\frac{\Mp^{}}{|\xis |}\, ,
\end{eqnarray}
where $\,\bar{\eta}_{12}^{}=(1-4M_\chi^2/\ucutt )^{\frac{1}{2}}_{}$\,
and the scattering energy takes the maximal value $\,E_{\max}^{}=\ucut\,$.\,
This shows that the perturbative unitarity bound requires the new scale
$\,\Mp/|\xis |\,$ to be higher than the scale of dark matter mass
$M_\chi^{}\,$.\footnote{Inspecting (\ref{SRNMCEFb}),
we see that the interactions involving more than $4\X$ need further expansion of $1/\Omega^2$,
which brings in additional $\xis/\Mp^2$ for each pair of $\X$. Thus the scattering channel
$\,\X \X \to \X \X\,$ places the best unitarity constraint on $\,\xis\,$.}\,
From \eqref{eq:PUBE0}, we further derive
\beqa
\label{eq:xih-xis-UB2}
\sqrt{|\xih\xis |\,} ~<~
\sqrt{\frac{4\pi}{\,3\,r\,\bar{\eta}_{12}^{}\,}\,}\frac{\,\Mp^{}\,}{\,\MX\,} \,,
\hspace*{9mm}
\eeqa
where $\,r = |\xis /\xih |\,$ and
$\,\bar{\eta}_{12}^{}=(1-4M_\chi^2/\ucutt )^{\frac{1}{2}}_{}$\,.\,
Different from \eqref{eq:xih-xis-UB},
we see that the bound \eqref{eq:xih-xis-UB2} is independent of the scattering energy $\,E$,\,
but inversely suppressed by the dark matter mass $\,\MX$.\,

\vspace*{1mm}

The second class of processes involves a pair of fermions,
$\,\phi\phi \to f\bar{f}$ or $\,\X\X \to f\bar{f}$\,.\,
At high energies, their amplitudes are dominated by $\,\OO(E^1)\,$ terms.
The amplitudes $\,\phi\phi \to f\bar{f}\,$ and $\,\X\X \to f\bar{f}\,$ are enhanced by
$\,\xih\,$ and $\,\xis\,$,\, respectively.
For $\,|\xih|,|\xis| \gg 1\,$,\, the unitarity bounds (\ref{eq:PUBE2}) and (\ref{eq:PUBE0})
from pure scalar scatterings are much stronger than these processes with a fermion pair.

\vspace*{4mm}
\section{\hspace*{-3.5mm} Analyzing Thermal Relic Density of GDM}
\label{sec:3}
\vspace*{2mm}

In this section, we study the property of the GDM $\,\X\,$ as a WIMP dark matter candidate.
We explore the intriguing possibility that the GDM alone fully accounts for the observed thermal relic density.
From this, we will analyze the viable parameter space for the GDM.
We find three independent parameters involved for this analysis:
the dark matter mass $\,\MX\,$ and two nonminimal couplings $(\xih, \,\xis)$.\,
As we will elaborate, in most of the parameter space, the prediction of thermal relic abundance
is only sensitive to the product of two nonminimal couplings $\,\xih\xis\,$.\,
Hence, our GDM construction is very economical and highly predictive.

\begin{figure}[h]
\vspace*{-3mm}
\centering%
\includegraphics[height=9cm,width=14.5cm]{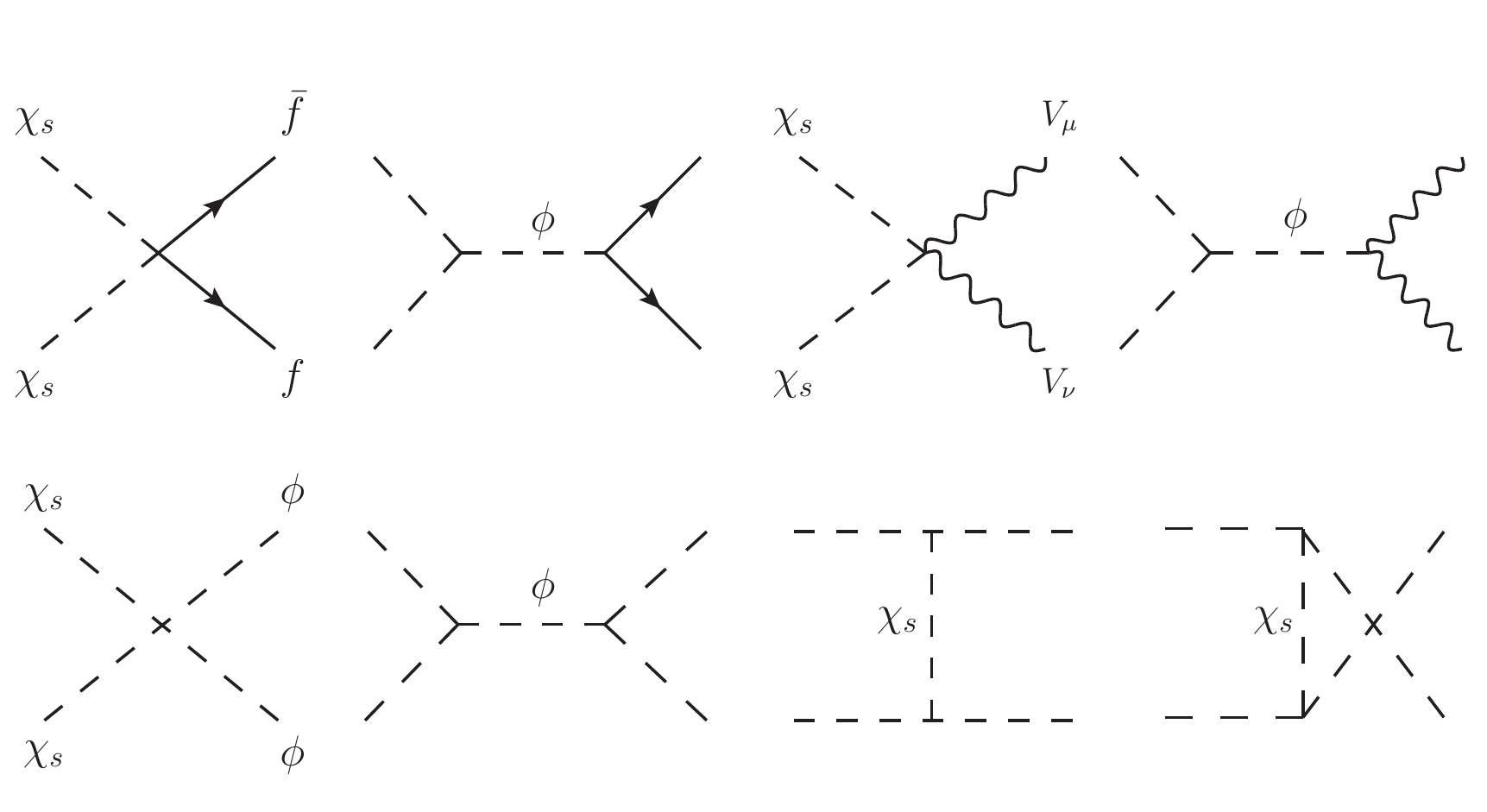}
\caption{The annihilation processes for the GDM, $\,\X\X\!\to VV,\,\phi\phi,\,f\bar{f}\,$.}
\label{fig:2}
\vspace*{4mm}
\end{figure}

In Fig.\,\ref{fig:2}, we display all channels for the GDM annihilations into the two-body final states
at leading order,\footnote{%
For channels with final states containing more two particles,
they may come from decays of off-shell heavy particles.
For instance, we can estimate the size of $\,\X\X\to W W^*\to W\bar{f}f'$.\,
In the intermediate mass-range $\,\fr{1}{2}m_W^{} < \MX < m_W^{}$,\,
the ratio $(\sigma_A v)_{Wff}^{}/(\sigma_A^{} v)_{bb}^{}\,$ is at most
$\,g^2m_W^2/(24\pi^2m_b^2) < 1\,$  for $\,\MX \lesssim m^{}_W$,\,
and further suppressed by the on-shell $W$ momentum for decreasing $\,\MX\,$.\,
Thus, it is reasonable to just count on the leading two-body annihilation channels.
Also, the loop-induced annihilations from the effective vertex \eqref{eq:FVAAXX}
are negligible.}\,
$\,\X\X\!\to VV,\,\phi\phi,\,f\bar{f}\,$,\,
where $\,V=(W,\,Z)\,$ and $\,f=(s,\, \mu,\, c,\, \tau,\, b,\, t)$.\,
Given the present sensitivities to WIMP via various experimental searches,
we will consider the GDM mass-range,
$\,\mathcal{O}(1\textrm{GeV})\lesssim \MX \lesssim \mathcal{O}(1\textrm{TeV})$.\,
From Eq.\,(\ref{eq:FVppXX}) and Eqs.\,(\ref{eq:FVVVXX})-(\ref{eq:FVqqXX}),
we see that the gravity-induced couplings to fermion and gauge boson are proportional
to their masses. Hence, for heavy mass-range
$\,\mathcal{O}(100\textrm{GeV})\lesssim M_\chi^{}\lesssim \mathcal{O}(1\textrm{TeV})$,\,
the dark matter annihilations are dominated by the channels
$\,\X\X \to W^+W^-, ZZ, \bar{t}t, \phi\phi$\,,\,
while for light mass-range $\,\MX\lesssim \mathcal{O}(10\textrm{GeV})$\,
only the annihilation channels
$\,\X\X\to \bar{b}b,\, \bar{c}c,\, \tau\tau,\,$ $\bar{s}s,\, \mu\mu$\,
are allowed at the freeze-out temperature.

\vspace*{1mm}

Before performing systematical numerical analyses,
we may first estimate the required size of the nonminimal couplings $(\xih,\,\xis)$
for accommodating the DM thermal relic density (when $\,\MX\,$ is away from any threshold or resonance).
For $s$-wave annihilation, the cross section dictated by the relic density abundance is
\begin{eqnarray}
\left<\sigma_A^{} v\right> \,\sim\,
3\times 10^{-26}\textrm{cm}^3\textrm{s}^{-1}
\,\sim\, 2.7\times 10^{-9} \textrm{GeV}^{-2}\, .
\end{eqnarray}
For $\MX$ much heavier than the weak scale,
the following heavy modes dominate in final states.
At leading order, the thermal averaged cross sections equal
the zero-temperature expression with $\,s\simeq 4M_\chi^2\,$,
\begin{eqnarray}
\langle\sigma_A^{} v\rangle_{VV}^{} \,\simeq\, \frac{9\xi_h^2\xi_s^2M_\chi^2}{\,\pi \Mp^4\,}, \quad~~
\langle\sigma_A^{} v\rangle_{tt}^{} \,\simeq\, \frac{27\xi_h^2\xi_s^2m_t^2}{\,\pi \Mp^4\,}, \quad~~
\langle\sigma_A^{} v\rangle_{\phi\phi}^{} \,\simeq\, \frac{9\xi_h^2\xi_s^2M_\chi^2}{\,\pi \Mp^4\,}\,,
\hspace*{10mm}
\end{eqnarray}
where we consider the parameter region of $\,|\xih|, |\xis| \gg 1$\,.\,
Note that the leading order contributions for gauge boson final states
come from the longitudinal modes and coincide with that of the Higgs final state.
For $\,\MX = \mathcal{O}(10^2-10^3)$GeV,
the product of nonminimal couplings is required to have a size around
$\,\sqrt{|\xih \xis|} \sim \OO(10^{14.5})$.\,
When $\,\MX\lesssim 100\,$GeV, the heavy final states
$\,WW, ZZ, hh, t\bar{t}\,$ decouple in advance,
and the cross section is dominated by the annihilations
$\,\X\X\to b\bar{b}, c\bar{c},\tau\tau$\,.

\vspace*{1mm}

To compute thermal relic density of dark matter with a wide mass range,
we will take into account threshold and resonance effects in the thermal integration\,\cite{Griest:1990kh}.
Compared with zero temperature case, annihilations into the final state slightly heavier than dark matter
could be active, due to its Boltzman distribution at finite temperature.
Also, when dark matter annihilate near the Higgs mass pole, i.e.,
$\,2\MX\sim m_\phi^{}$,\, the cross section is largely enhanced over the case away from the pole.
To properly treat the cross section around the pole,
we will make thermal integration numerically without any expansion for velocity.
Using the couplings of (\ref{eq:FVppXX}) and (\ref{eq:FVVVXX})-(\ref{eq:FVqqXX}),
we derive zero temperature annihilation cross section for the relevant final states,
\beqs
\begin{eqnarray}
\label{eq:sAvSMT0}
(\sigma_A v)_{ff}^{} & \simeq& \frac{N_c m_f^2}{\,4\pi \Mp^4\,}
\frac{(24\xih\xis M_\chi^2)^2}{\,(4M_\chi^2\!-m_\phi^2)^2\!+\! m_\phi^2\Gamma_\phi^2\,}
\left(\!1-\frac{m_f^2}{M_\chi^2}\right)^{\!\!\frac 3 2} \!,
\\[2mm]
(\sigma_A v)_{VV}^{} & \simeq & \frac{\delta_V m_V^4}{\,16\pi M_\chi^2\Mp^4\,}
\frac{(24\xih\xis M_\chi^2)^2}{\,(4M_\chi^2\!-m_\phi^2)^2\!+\!m_\phi^2\Gamma_\phi^2\,}
\left(\!1-\frac{m_V^2}{M_\chi^2}\right)^{\!\!\frac 1 2}
\!\left[2+\left(\!1-2\frac{M_\chi^2}{m_V^2}\right)^{\!\!2}\right] \!,
\\[2mm]
\label{eq:sAv-hh}
(\sigma_A v)_{\phi\phi}^{} & \simeq &
\frac{(24\xih\xis M_\chi^2)^2}{\,64\pi M_\chi^2\Mp^4\,}
\left(\!1+\frac{3m_\phi^2}{\,4M_\chi^2\!-\!m_\phi^2\,}
- \frac{3\xih\xis \vew^2 m_\phi^4}{\,\Mp^2M_\chi^2(2M_\chi^2-m_\phi^2)\,}\right)^{\!\!2}
\left(\!1-\frac{m_\phi^2}{M_\chi^2}\right)^{\!\!\frac 1 2}\!,
\hspace*{13mm}
\end{eqnarray}
\eeqs
where $\,f=t, b, c, s, \tau, \mu\,$ and $\,N_c^{}=3\,(1)\,$ for quarks (leptons).
We also denote $\,V=W,Z\,$ and $\,(\delta_W^{},\,\delta_Z^{})=(2,\,1)$.\,
For the parameter range of interest, we only keep the leading order contributions
under $\,|\xih|, |\xis|\gg 1\,$.\, Thus, the cross sections are controlled by the DM mass $\,\MX\,$
and the product of nonminimal couplings $\,\xih\xis$\,.\,
The width of a SM Higgs boson with mass 125\,GeV is rather small,
$\,\Gamma_\phi^{\text{SM}}\simeq 4.03\,$MeV.\,
In our model, the Higgs total width $\Gamma_\phi^{}$ could deviate from the SM value
only when the invisible decay channel $\,\phi\to\X\X\,$ is open.
Since the DM mass-range for active annihilation process $\,\chi\chi\to\phi\phi\,$
is far away from resonance region, we can safely neglect $\Gamma_\phi^{}$
in Eq.\,\eqref{eq:sAv-hh}.
We present the calculation of thermal averaged cross sections by including the threshold and resonance effects
in Appendix\,\ref{app:TRDdetail}.

\vspace*{1mm}

Given the thermal average cross section $\,\langle\sigma_A^{}v\rangle$\, as function of the DM mass $\,\MX\,$
and coupling product $\,\xih\xis\,$,\, we will derive thermal relic abundance.
It is convenient to define a ratio $\,x\equiv \MX/T\,$.\,
Thus, the freeze-out temperature $\,\xfo^{}=\MX /T\fo^{}\,$
can be derived from the following formula to a good accuracy \cite{Kolb:1990vq},
\beqs
\begin{eqnarray}
\hspace*{-10mm}
&& \frac{\xfo^2}{\,(2+c)\tilde{\lambda}\,\langle\sigma_A^{}v\rangle\fo^{}\,}
\,\simeq\, c\,a\,\xfo^{3/2}e^{-\xfo^{}},
\\[2mm]
\hspace*{-10mm}
&& \tilde\lambda \,\equiv\,\frac{\,2\sqrt{2}\pi\,}{\,3\sqrt{5}\,}
\frac{g_{*S}^{}}{\sqrt{g_*^{}}}\MX M_\textrm{Pl}^{} \,,
\quad~~
a \,\equiv\, \frac{45}{\,2\pi^2(2\pi)^{3/2}g_{*S}^{}\,}\, ,
\end{eqnarray}
\eeqs
where $\,\langle\sigma_A^{}v\rangle\fo^{}\,$
denotes the thermal averaged cross section at the freeze-out temperature $\,T\fo^{}\,$.\,
We denote the total effective relativistic degrees of freedom as $\,g_*^{}$\,
and its counterpart for entropy as $\,g_{*S}^{}\,$.\,
We further assume that all species in the universe have the same temperature,
and $\,g_*^{}\simeq g_{*S}^{}$.\,
The coefficient $\,c\,$ is a free-parameter for fitting the numerical solution,
and we use the conventional choice: $\,c(c+2)=1\,$ \cite{Kolb:1990vq}.
Then, above equation is simplified as
\begin{eqnarray}\label{eq:xfG}
\xfo^{} ~\simeq~
\ln \frac{\,0.19\MX\Mp \langle\sigma_A^{}v\rangle\fo^{}\,}{\sqrt{g_*^{}}\sqrt{\xfo^{}}} \,.
\end{eqnarray}
Integrating out the differential equation for number density per comoving volume,
we can derive thermal relic abundance,
\begin{eqnarray}
\label{eq:DMRDG}
\Omega_{\chi 0}^{}h^2 \,=\,
\frac{\,h^2 n_\chi^{}\MX\,}{\rho_c}  \,\simeq\,
\frac{\,2.12\!\times\!10^8\,\textrm{GeV}^{-1}\,}{\,\sqrt{g_*^{}}J(\xfo^{})M_\textrm{Pl}^{}\,}\,,
\quad~~~
J(\xfo^{}) \,\equiv\, \int_{\xfo}^\infty \!\!dx\,\frac{\langle\sigma_A^{}v\rangle}{x^2} \,, ~~~~
\end{eqnarray}
where $\,\rho_c^{}/h^2=1.88\!\times\! 10^{-29}\textrm{g cm}^{-3}$,\, and
around the freeze-out, $\,g_*^{}\simeq g_{*S}^{}\simeq 106+g_\chi^{}\,$.\,
In context of $\Lambda$CDM scenario, the latest measurement from Planck satellite gives,
$\,\Omega_{\chi 0}^{}h^2=0.1199\pm 0.0027$ \cite{Planck2013}.
In the minimum setup, the relic abundance is only sensitive to $\,\MX\,$ and $\,\xih\xis\,$.\,

\begin{figure}[t]
\label{fig:3}
\centering%
\includegraphics[height=7cm,width=7.6cm]{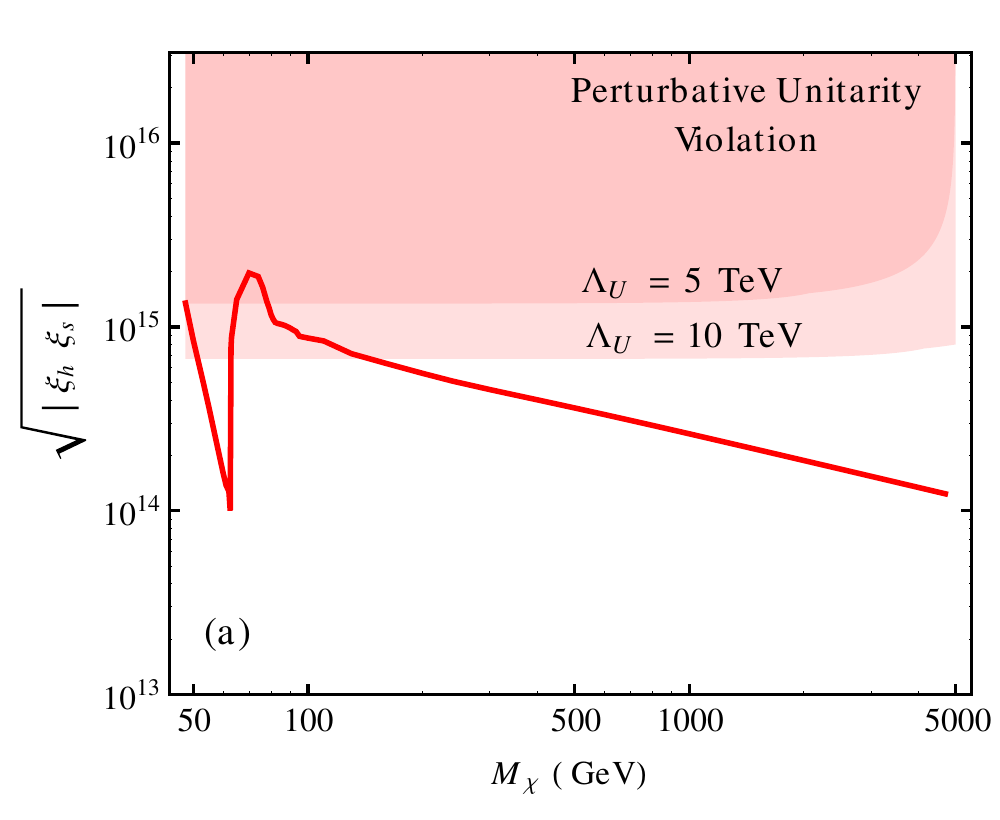}\,
\includegraphics[height=7cm,width=7.6cm]{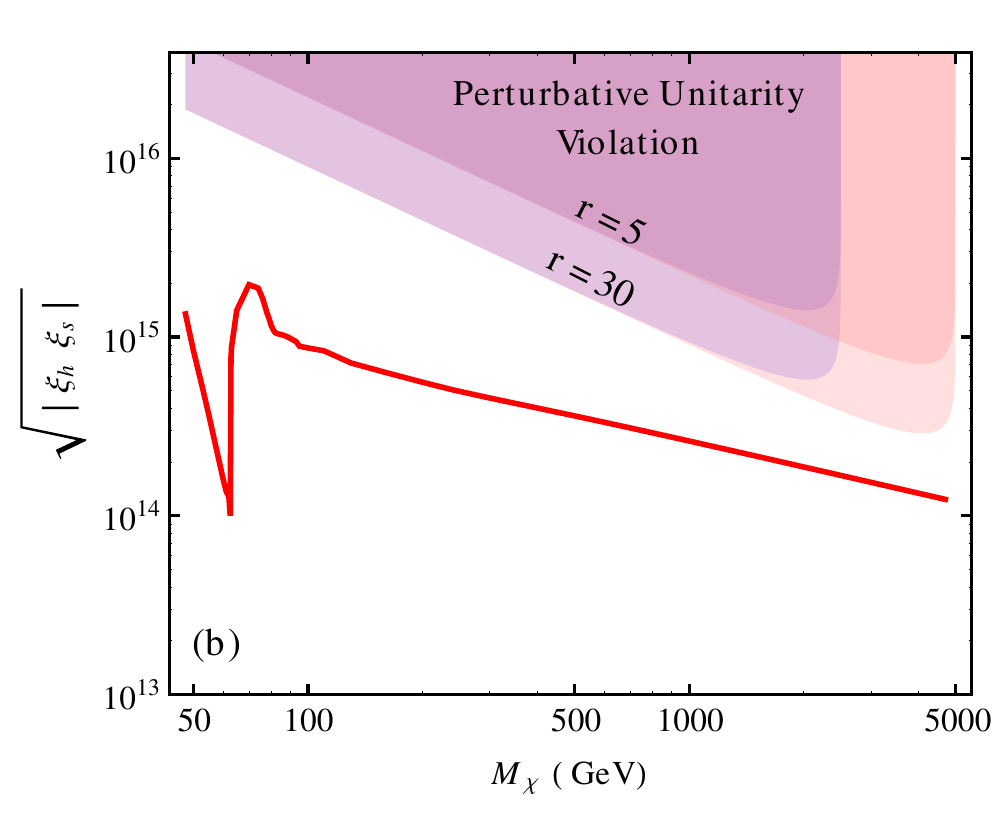}
\vspace*{-5mm}
\caption{Viable parameter space in $\,\MX-\sqrt{|\xih\xis|}\,$ plane.
In both plots, the red solid curve is predicted by generating the observed
thermal relic density $\,\Omega_{\chi,0}^{}h^2 \simeq 0.12$\,.\,
In plot-(a), the shaded pink areas present the perturbative unitarity violation regions
from condition \eqref{eq:xih-xis-UB}, where we input a typical coupling ratio
$\,r=|\xis/\xih |=5\,$ and set the sample UV cutoff $\,\ucut =5,\,10\,$TeV.
In plot-(b), the shaded pink (purple) areas present the perturbative unitarity violation regions
from condition \eqref{eq:xih-xis-UB2} with $\,\ucut =10\,(5)\,$TeV, where we input the typical
coupling ratios $\,r=5,\,30$,\, respectively.}
\vspace*{3mm}
\end{figure}

In Fig.\,\ref{fig:3}, we present the contours of thermal relic density in
$\,\MX-\sqrt{|\xih\xis|}$\, plane, where the red solid curve corresponds to
$\,\Omega_{\chi 0}^{}h^2\simeq 0.12$\,.\,
For illustration, we focus on the GDM mass range above $\,\MX \simeq 48\,$GeV.\,
This is the lower bound set by the Higgs invisible decay constraint (cf.\ Fig.\,\ref{fig:6}),
above which the narrow width assumption of Higgs holds well.
With the increase of $\,\MX\,$,\, more heavy SM modes contribute to the annihilation cross sections,
and thus the required $\,\sqrt{|\xih\xis |}\,$ becomes smaller.
Around the Higgs mass-pole, $\,2\MX\simeq m_\phi^{}$,\,
the required $\,\sqrt{|\xih\xis |}\,$ becomes almost one order of magnitude smaller
due to the resonance enhancement.
For $\,\MX \lesssim m_W^{}$,\,
the forbidden channel $\,\X\X\to WW\,$ contributes right below the threshold
due to the thermal fluctuations shown in Appendix\,\ref{app:TRDdetail}.
Thus, the required nonminimal coupling for realizing the relic density could be
smaller than what expected for zero temperature case.
From Eq.\,\eqref{eq:DMRDG}, we note that $\,\Omega_{\chi0}^{}h^2$\,
is roughly proportional to $\,(\xih\xis)^{-2}$\, via the thermal averaged cross section,
i.e., $\,\sqrt{|\xih\xis|}\propto (\Omega_{\chi 0}^{}h^2)^{-1/4}$.\,
This means that the solid curve in Fig.\,\ref{fig:3}
is very insensitive to the experimental uncertainty of
$\,\Omega_{\chi 0}^{}h^2\,$.\,
Hence, provided that $\,\X\,$ dominates in thermal relics,
the constraint on $\,\sqrt{|\xih\xis |}\,$ is so robust that
almost no visible variations from the solid curve are allowed.

\vspace*{1mm}

The present analysis is fully based upon perturbative expansion in the effective theory
formulation \cite{EFT}, which combines the SM with nonrenormalizable Einstein general relativity.
Such an effective theory normally has an UV cutoff scale $\,\ucut\,$,
above which the perturbative expansion breaks down
and new physics is expected to show up.\footnote{%
Our current effective theory study considers TeV scale quantum gravity
with UV cutoff $\,\ucut =\order{10\text{TeV}}\,$.\,
We are not concerned with any detail of the UV dynamics above $\,\ucut\,$.\,
Many well-motivated TeV scale quantum gravity theories exist on the market.
For instance, an extra dimensional model with compactification scale of
$\,\order{10\text{TeV}}\,$ will reveal its Kaluza-Klein modes at energies above this scale.}
For the validity of our perturbative analysis,
we should derive perturbative unitarity constraints on the parameter space.
The shaded areas in the two plots of Fig.\,\ref{fig:3} depict the unitarity violation regions
from two different unitarity conditions \eqref{eq:xih-xis-UB} and \eqref{eq:xih-xis-UB2}.
The first condition \eqref{eq:xih-xis-UB} is derived from $\OO(E^2)$
leading terms of the scattering amplitudes.
It is insensitive to the DM mass $\,\MX$,\, as shown in Fig.\,\ref{fig:3}(a).
It is also insensitive to the coupling ratio $\,r=|\xis/\xih|\,$ for $\,r\gtrsim 5\,$.\,
So we will take a sample input of $\,r=5\,$ for illustration.
We present the unitarity violation regions in Fig.\,\ref{fig:3}(a),
for the sample UV cutoff (set by unitarity bound) $\,\ucut =5,\,10$\,TeV,\, respectively.
We see that the unitarity constraint is mild, and our effective theory remains perturbative
for a wide range of $\,\MX \,$.\,
The second unitarity condition \eqref{eq:xih-xis-UB2} is derived from
$\,\X\X\to\X\X\,$ channel at $\,\OO(E^0)\,$.\,
This is shown in Fig.\,\ref{fig:3}(b),
where the bounds are quickly enhanced with the increase of DM mass.
Here, the shaded purple and pink regions correspond to the
sample cutoff scale $\,\ucut = 5,\,10\,$TeV,\,
respectively. In each case, we have presented the unitarity constraints
for the coupling ratio $\,r=5\,$ (darker shaded area) and
$\,r=30\,$ (lighter shaded area).\,
This plot shows that for a valid perturbation analysis, the ratio $\,r\,$
cannot be too large, namely, the values
of the two nonminimal couplings $\xis$ and $\xih$ should not have a large hierarchy.
Finally, we note that the current LHC measurements on the Higgs signal rates
put a mild constraint on the coupling $\,\xih$\,
via the kinetic rescaling factor $\,\zeta\,$ in \eqref{eq:zeta}.
For instance, from the latest CMS (ATLAS) data \cite{LHC2014},
we can infer the $3\sigma$ upper bound, 
$\,|\xih | < 3.4\times 10^{15}\,$ ($\,|\xih | < 2.3\times 10^{15}\,$).
In the following analysis,
we will study various experimental searches of GDM within the viable parameter space
that generates the DM thermal relic abundance
and obeys the perturbative unitarity bounds (Fig.\,\ref{fig:3})
as well as the current LHC bound on the nonminimal coupling $\,\xih\,$.

\vspace*{1mm}

As a final remark in this section, we note that even though the GDM can have a large
nonminimal coupling $\,\xis\,$,\,
it would not cause any large effect on the long distance gravitational behavior.
Since GDM field has no VEV, its nonminimal coupling \eqref{eq:NMC} in Jordan frame
does not contribute to the Planck mass.
This nonminimal coupling term is fully transformed away in Einstein frame,
and is replaced by a set of higher dimensional operators involving effective interactions
between the GDM and SM fields (Sec.\,\ref{sec:2.2}).
These dimension-6 operators are suppressed by
$\,(\xi_s^2,\,\xi_h^2,\,\xis\xih )(\vew^2,E^2)/\Mp^2\,$
and do not cause any sizable effect at long distance (low energy).
They are further constrained by perturbative unitarity bounds
(Sec.\,\ref{sec:2.3} and Fig.\,\ref{fig:3}) at high energies.
These GDM effective couplings properly generate the observed DM relic density.
It means that our GDM belongs to a kind of WIMP dark matter
and does not cause extra visible change at long distance.

\vspace*{4mm}
\section{\hspace*{-3.5mm} GDM Detections and Collider Searches}
\label{sec:4}
\vspace*{2mm}

In this section, we explore various searches of the GDM.
With the gravity-induced interactions between the GDM and SM particles,
we find it is quite difficult to probe the GDM by direct detections due to
the small-momentum suppression. On the other hand, indirect detections can be promising
to reach the parameter space that successfully accounts for thermal relic abundance.
Finally, we study the collider searches of the GDM from the Higgs invisible decay,
and further discuss the probe of a heavier GDM $\X$ at the LHC\,(14\,TeV)
and future high energy hadron colliders.

\vspace*{3.5mm}
\subsection{\hspace*{-3.5mm} Direct Detection of GDM}
\vspace*{2mm}

\begin{figure}[t]
\centering%
\includegraphics[height=8.6cm,width=10.2cm]{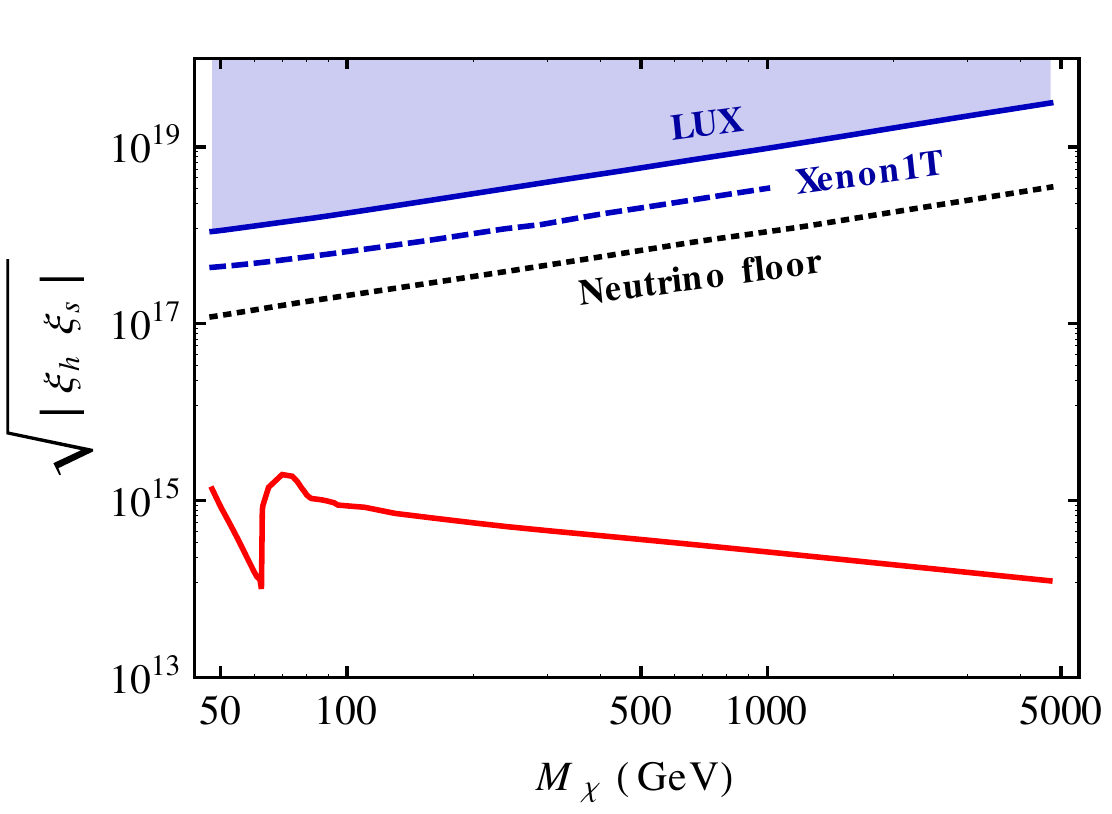}
\caption{Exclusions in $\,\MX \!-\!\sqrt{|\xih\xis|}$\, plane at 90\%\,C.L., which are derived
from measuring spin-independent GDM-nucleon cross sections via direct detection experiments.
The shaded regions are excluded, and the blue dashed curve gives the reach (upper bound) of
future experimental projection. The region below black short-dashed curve denotes the parameter space
sensitive to neutrino background. The red solid curve is our prediction which accounts for the
DM thermal relic abundance $\Omega_{\chi 0}^{}h^2 = 0.12$\,.}
\label{fig:4}
\end{figure}

In the present model, the GDM scalar $\,\X\,$ interacts with nucleons via the gravity-induced
interactions between light fermions and $\,\X\,$.\,
As a scalar dark matter, the GDM-nucleon interaction is spin-independent.
From Eq.\,(\ref{eq:FVqqXX}), we derive the GDM-nucleon scattering cross section
under the limit $\,|\xis|,|\xih| \gg 1\,$,
\begin{eqnarray}
\label{eq:DDSIxs}
\sigma_{\textrm{SI}}^{} ~\simeq~
\frac{f_N^2m_N^4}{\,4\pi(M_\chi\!+\! m_N)^2\Mp^4\,}
\left(\xis+\frac{\,6\xih\xis t\,}{\,t\!-\! m_\phi^2\,}\right)^{\!\!2}  ,
\end{eqnarray}
where $\,f_N^{}\,$ is the effective form factor, which
can be estimated from the QCD chiral perturbative theory, the pion-nucleon scattering
and the lattice simulations.
We will use $\,f_N^{}=0.345$\, for the following estimate \cite{fN}.
For the nucleon mass, we input the averaged mass of proton and neutron,
$\,m_N^{}=0.939\,$GeV.\,
The typical scale of momentum-exchange is around $\,100\,\textrm{MeV}$,\, i.e.,
$\,t\approx -(100\,\textrm{MeV})^2$.\,
In the parameter space of interest, the second term in the parentheses of Eq.\,(\ref{eq:DDSIxs})
is dominant.\footnote{This approximation numerically holds well for $\,\xih\,$ within
the perturbatively unitary range.}\,
Comparing with annihilation cross section at $\,\sqrt{s}\simeq 2\MX\,$,\,
we see that the GDM-nucleon cross section has a momentum-suppression factor $\,(t/m_\phi)^2$\,,\,
due to the energy-dependent structure of $\,\X-\X-\phi$\, vertex.

\vspace*{1mm}

In Fig.\,\ref{fig:4}, using the data of relevant direct detection experiments,
we present a summary of their exclusions (at 90\%\,C.L.) in $\,\MX -\sqrt{|\xih\xis|}$ plane.
For $\,\MX \gtrsim 10\,$GeV, the strongest constraint comes from LUX experiment\,\cite{LUX2013},
as depicted by blue solid curve.
The future reach of Xenon1T projection\,\cite{XENON1T} is represented by the blue dashed curve.
Due to the low-momentum suppression of $\,\sigma_{\textrm{SI}}^{}\,$ in Eq.\,(\ref{eq:DDSIxs}),
the magnitude of $\,\sqrt{|\xih\xis|}$\, as dictated by the observed DM thermal relic abundance
(red solid curve) is even below the required sensitivity to the neutrino background and
out of the reach of direct detection.
Hence, the GDM that can fully account for the thermal relic abundance
is unlikely to be detected through the nucleus recoil.
This is a typical feature of the GDM in contrast to other WIMP DM candidates.

\vspace*{3.5mm}
\subsection{\hspace*{-3.5mm} Indirect Detection of GDM}
\vspace*{2mm}

Many astrophysical experiments aim at finding indirect evidences of dark matter annihilations in the sky.
In the present model, we find that searching for gamma ray signals from target with high dark matter density
is most promising.\footnote{PAMELA\,\cite{PAMELA} and AMS02\,\cite{AMS02} reported cosmic ray electron-positron
excess recently, which may be explained by DM annihilations or astrophysical sources (such as quasar).
In the present model, since GDM interacts with SM particles via gravity
and the interaction strength is proportional to the SM particles masses,
the GDM coupling to electrons is too small to account for this excess.
The measurements of cosmic ray antiproton, which could be produced from hadronization of the primary products of
DM annihilations, may serve as another way for DM indirect detection. But, the interpretation suffers
larger uncertainty from modeling of the antiproton propagation in galaxies \cite{antiP}.}\,

\vspace*{1mm}

There are two types of gamma ray signals. One is monochromatic photon ``line'' arising from
the dark matter annihilation $\,\X\X\to\gamma X\,$,\,
where $\,X\,$ denotes any other possible SM bosons.
The other one is a diffuse continuum spectrum from secondary production of photons from primary dark matter
annihilation $\,\X\X \to W^+W^-, ZZ,\, b\bar{b},\, \tau^+\tau^-,\, \mu^+\mu^-$.\,
The secondary photon may be initiated from final state radiation or hadronization
with decays $\,\pi^0\to\gamma\gamma$\,.\,
In the following, we study the impacts of these measurements on our model in turn.

\vspace*{1mm}

As discussed in Sec.\,\ref{sec:2.2}, the effective operators for
$\,\X\X \to \gamma X\,$ can be induced from gravitational interactions
at one-loop order. From (\ref{eq:FVAAXX}), we infer the zero temperature cross section
for $\,\X\X\to\gamma\gamma$\, and $\,\X\X\to\gamma Z$\,,\,
\beqs
\begin{eqnarray}
(\sigma_A v)_{\gamma\gamma}^{} &\,=\,&
\left(\frac{\alpha}{\pi}\right)^2\frac{\,16M_\chi^2\,}{\pi \Mp^4}
\left|\mathcal{C}_\gamma\right|^2
\frac{(24\xih\xis M_\chi^2)^2}{(4M_\chi^2-m_\phi^2)^2+m_\phi^2\Gamma_\phi^2} \,,
\\[2mm]
(\sigma_A v)_{\gamma Z}^{} &=&
\left(\frac{\alpha}{\pi}\right)^2\frac{8M_\chi^2}{\,\pi\Mp^4\,}
\left|\mathcal{C}_{\gamma z}\right|^2\frac{(24\xih\xis M_\chi^2)^2}
{\,(4M_\chi^2\!-\!m_\phi^2)^2\!+m_\phi^2\Gamma_\phi^2\,}
\left(\!1-\frac{m_Z^2}{4M_\chi^2}\right)^{\!\!3} ,
\end{eqnarray}
\eeqs
where $\,\alpha\simeq 1/128$\, is the fine structure constant,
and $\,\mathcal{C}_\gamma^{}, \mathcal{C}_{\gamma z}^{}\,$ are energy dependent loop-factors defined
in (\ref{eq:loopfactors}).
The annihilation channel $\,\X\X\to\gamma Z\,$ is active for $\,\MX >m_Z^{}$\,.\,
In comparison with tree-level processes, these two channels are suppressed by a loop factor.
Nevertheless, since the ``line" shape search features a better sensitivity than that of the continuum
spectrum, the constraint from monochromatic spectrum would be potentially important.
Provided that the DM annihilations into $\,\gamma\gamma$\, and \,$\gamma Z$\, are the only sources
to generate gamma ray line, it is possible to extract an upper bound on the quantity
$\,2(\sigma_A^{}v)_{\gamma\gamma}^{}+(\sigma_A^{}v)_{\gamma Z}^{}$\,
from galactic center $\gamma$-ray line search \cite{Fedderke:2013pbc},
i.e., Fermi-LAT in low photon energy range\,\cite{FermiLATline} and H.E.S.S in high energy range\,\cite{HESSline}.
The limits depend on the DM halo profiles as well as the signal region of interest (selected by
the experimental group for analyses).

\begin{figure}
\centering%
    \includegraphics[height=8cm,width=10cm]{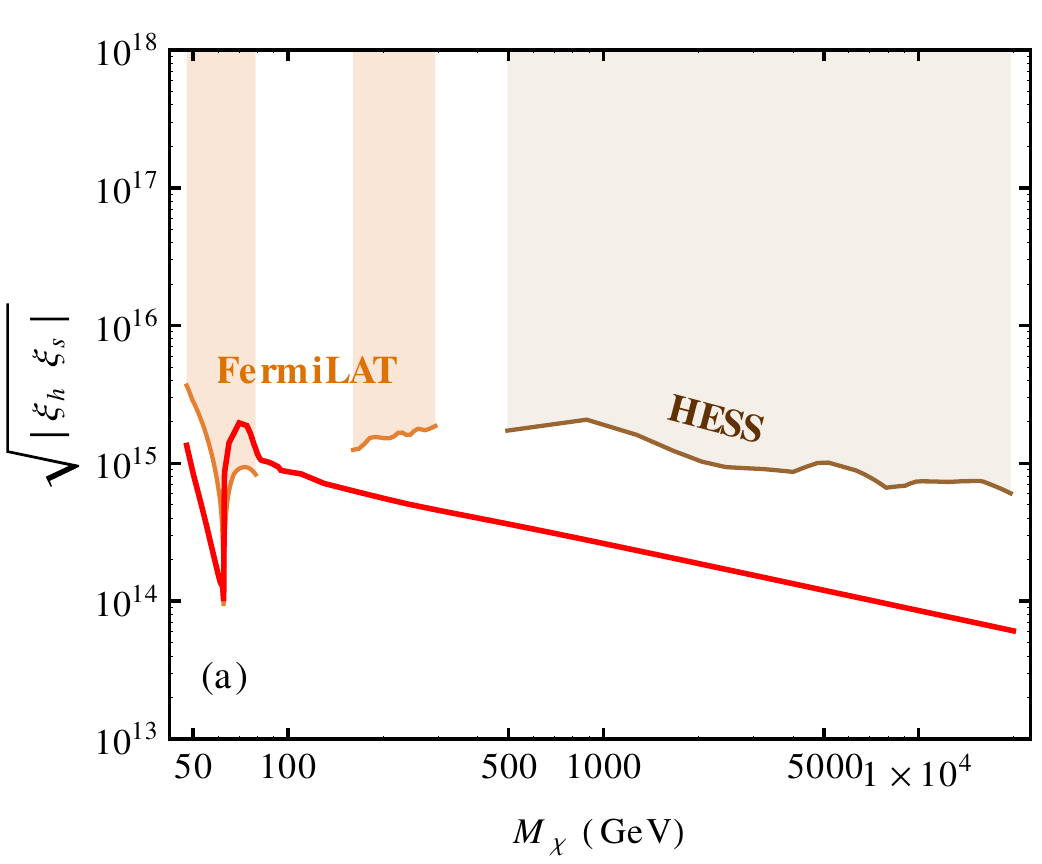}\\
    \includegraphics[height=8cm,width=10.5cm]{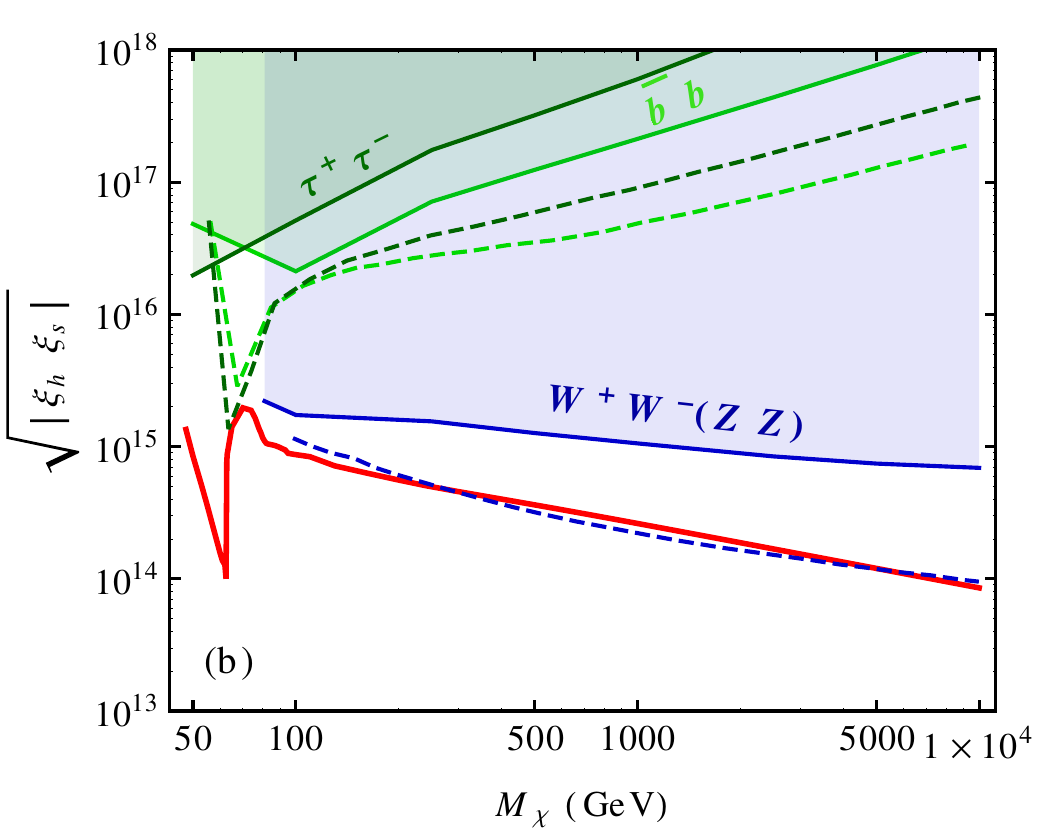}
\vspace*{-2mm}
\caption{Constraints in $\,\MX-\sqrt{|\xih\xis|}$\, plane from indirect DM detections.
(a).\ 95\%\,C.L.\, exclusions from gamma ray line search. The light brown curve depicts
the strongest limit from FermiLAT in the low photon energy region, and the dark brown curve
denotes the limit from H.E.S.S\,\cite{Fedderke:2013pbc}. The areas above these curves are excluded.
(b).\ Exclusions from gamma ray continuum spectrum. The shaded regions are excluded at 95\%\,C.L.
The blue, light green and dark green curves represent the bounds from detections via
three primary annihilation channels $\,W^+W^- (ZZ),\, b\bar{b}\,$ and $\,\tau^+\tau^-$,\, respectively.
The solid and dashed curves denote bounds from Fermi-LAT and CAT (projection), respectively.
In each plot, the red solid curve gives the prediction by realizing the GDM thermal relic density
$\,\Omega_{\chi 0}^{}h^2 = 0.12$\,.}
\label{fig:5}
\end{figure}

\vspace*{1mm}

To demonstrate the potential of these experiments for testing
our model, we present the strongest constraint from
the gamma-ray line search\,\cite{Fedderke:2013pbc} for illustration.
Fig.\,\ref{fig:5}(a) depicts these constraints
in $\,\MX\!-\!\sqrt{|\xih\xis |}$\, plane,
where the shaded regions are excluded at 95\%\,C.L. The light brown curve is extracted
from the searches of FermiLAT \cite{FermiLATline}.
In the intermediate mass range $80-160$\,GeV,
since the line shape is sensitive to relative strength of the two processes,
no reliable model-independent limit could be inferred \cite{Fedderke:2013pbc}.
The dark brown curve is extracted from H.E.S.S \cite{HESSline}.
As before, the red solid curve is our GDM prediction
by accommodating the DM thermal relic abundance.
We see that the GDM with mass between $\,60-80\,$GeV is already excluded by FermiLAT.
In low mass range below $\,m_\phi^{}/2$,\,
due to the resonance enhancement from thermal integration, i.e.,
$\,(\sigma_A^{}v)\lesssim \langle\sigma_A^{}v\rangle$,\, the GDM prediction is still viable.
For the GDM mass $\,\MX >80$\,GeV, the relic density
is dominated by the tree-level annihilation into heavier final states. In this mass range,
our prediction is significantly below the reach of the gamma ray ``line" searches.

\vspace*{1mm}

Next, we study constraints on dark matter annihilation cross sections
from diffuse continuum spectrum. The latest results come from the 4-years data of Fermi-LAT observation
of 15 Milky Way dwarf spheroidal satellite galaxies\,\cite{FermiLAT2013}.
In the future, the next generation experiments with better angular resolution (such as CTA \cite{CTA})
will largely improve the sensitivity over a wider mass range.
Normally, the upper limit on DM annihilation cross sections is extracted
by assuming 100\% branching fraction for each primary annihilation channel.
Since these limits are sensitive to the spectrum shape for each channel,
they could not be straightforwardly mapped to a given model where all annihilation channels contribute
in a certain pattern. Nevertheless, following Ref.\,\cite{Fedderke:2013pbc},
we may estimate the conservative constraint by taking into account the fraction of each channel
in the total annihilation cross section. The bound is derived as follows,
\begin{eqnarray}
\label{eq:BRID}
(\sigma_A^{}v)^{95\%,\textrm{res}}_{jj}\,=~
\frac{(\sigma_A^{}v)^{95\%}_{jj}}{\textrm{BR}_{jj}^{}}\, ,
\end{eqnarray}
where $\,(\sigma_A^{}v)^{95\%}_{jj}\,$ is the experimental upper bound.
$\,\textrm{BR}_{jj}^{} \equiv (\sigma_A^{}v)_{ii}^{}/(\sigma_A^{}v)_{\textrm{tot}}^{}$,\,
with $\,(\sigma_A^{}v)_{jj}^{}$\,
defined in Eq.\,(\ref{eq:sAvSMT0}) and $(\sigma_A^{}v)_{\textrm{tot}}^{}$
summing over all these channels.
Note that $\,\textrm{BR}_{jj}^{}$\, is only a function of mass $\,\MX\,$,\,
and is insensitive to $(\xih,\,\xis )$.\,
We then deduce the lower bound on $\,\sqrt{|\xih\xis|}$\,
from $\,(\sigma_A^{}v)^{95\%,\textrm{res}}_{jj}$\,
for $\,W^+W^-(ZZ),\, b\bar{b},\, \tau^+\tau^-$,\,
respectively.\footnote{Since there is no distinction between $W^+W^-$ and $ZZ$
in view of secondary gamma ray spectrum, the first mode $\,W^+W^-(ZZ)$\, corresponds to the sum $\,(\sigma_A^{}v)_{WW}^{}+(\sigma_A^{}v)_{ZZ}^{}$.}\,
The constraints are summarized in Fig.\,\ref{fig:5}(b),
where the blue, light green and dark green curves represent three primary annihilation channels
\,$W^+W^- (ZZ),\, b\bar{b}$\, and $\,\tau^+\tau^-$,\, respectively.
The shaded regions above solid curves are excluded by Fermi-LAT experiment at $95\%$\,C.L.,
and the dash curves present the sensitivity of CAT.
In the mass range $\,\MX \gtrsim 100\,$GeV, the strongest constraint on our model
comes from measurements of $\,W^+W^-(ZZ)$\, channels.
Our prediction from the GDM thermal relic abundance is within reach of
the future indirect detection experiments.
Recently, some studies suggested the gamma ray excess from Galactic Center \cite{Daylan:2014rsa},
which can be interpreted as a signal predicted by a $\,31-40\,$GeV dark matter annihilating
mostly into $\,b\bar{b}\,$ final state with cross section
$\,(\sigma_A^{}v)=(1.4-2.0)\!\times\! 10^{-26}\,\textrm{cm}^3\textrm{s}^{-1}$.\,
In the present model, the dominant annihilation in this intermediate mass range is indeed the
$\,b\bar{b}\,$ channel, but the required parameter range in $\,\MX\!-\!\sqrt{|\xih\xis|}$\, plane
is already excluded by Higgs invisible decays.

\vspace*{3.5mm}
\subsection{\hspace*{-3.5mm} Collider Searches for GDM}
\vspace*{2mm}

\begin{figure}[t]
\centering%
\includegraphics[height=8.5cm,width=11.9cm]{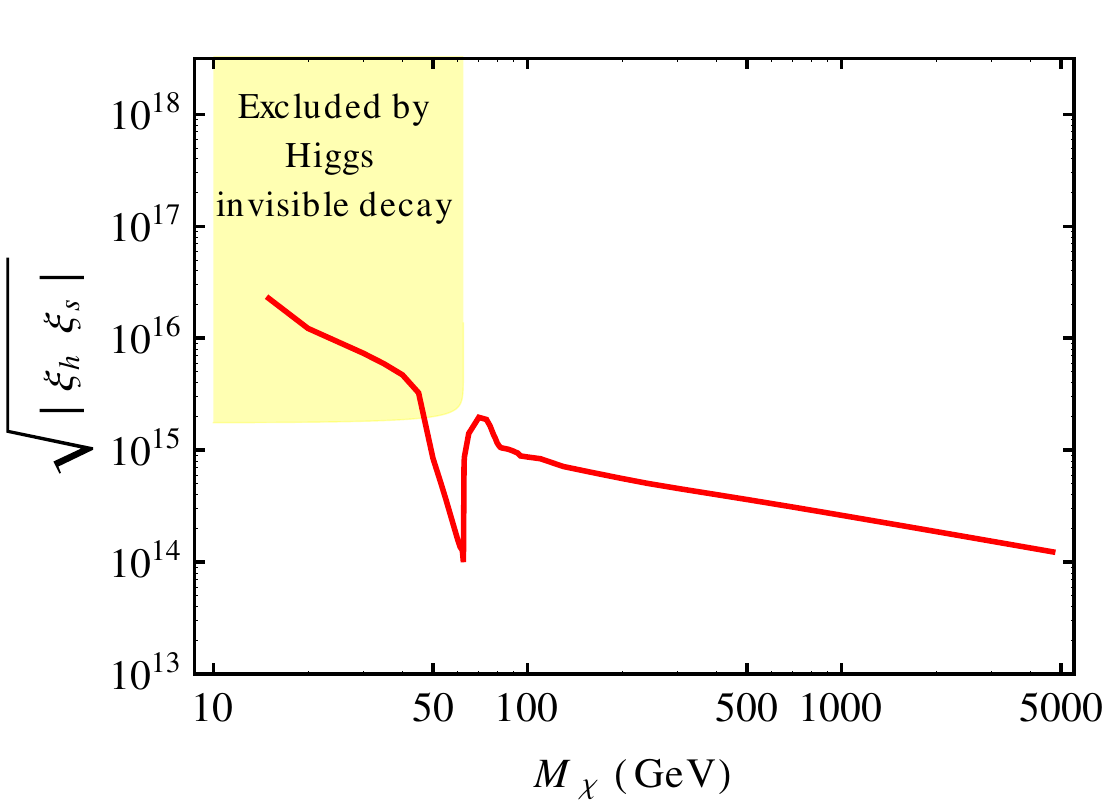}
\caption{Constraint from searching for Higgs invisible decays at the LHC \cite{Hinv}.
The yellow region is excluded at $95\%$\,C.L.
The red solid curve presents our prediction by generating the observed DM thermal relic density
$\,\Omega_{\chi 0}^{}h^2 = 0.12$\,.}
\label{fig:6}
\end{figure}

The GDM may be produced at hadron colliders in several ways.
For a light GDM with mass $\,\MX < m_\phi^{}/2$\,,\,
it can be produced via invisible decays of the 125\,GeV Higgs boson,  $\,\phi\to\X\X\,$,\,
due to the cubic vertex $\,\X-\X-\phi\,$ in (\ref{eq:FVsXX}).
For $\,|\xih|,|\xis|\gg 1$\,,\, we deduce the invisible decay width,
\begin{eqnarray}
\label{eq:width-H-XX}
\Gamma(\phi\to\!\X\X)
~=~ \frac{\,(3\xih\xis \vew)^2m_\phi^3\,}{8\pi\Mp^4}
\sqrt{1-\left(\!\frac{2M_\chi}{m_\phi}\!\right)^{\!\!2}\,} \, .
\end{eqnarray}
Accordingly, its invisible decay branching fraction is given by
\beqa
\textrm{BR}_{\chi\chi}^{} =\,
\frac{\Gamma(\phi\!\to\!\X\X)}{\,\big[\,\Gamma_\phi^{\textrm{SM}}+\Gamma(\phi\!\to\!\X\X)\big]\,},
\eeqa
where $\,\Gamma_\phi^{\textrm{SM}}\simeq 4.3\,$MeV\,
denotes the Higgs decay width from the SM contributions alone.
Currently, LHC searches invisible Higgs decays via the vector boson associated production,
vector boson fusion, and top associated production.
By assuming the SM production rate, the best upper limit on the invisible branching fraction
comes from combining all existing measurements at the LHC,
$\,\textrm{BR}_{\text{inv}}^{} \! < 40\%$\, at $95\%$\,C.L.\,\cite{Hinv}.
Setting $\,\textrm{BR}_{\text{inv}}^{}=\textrm{BR}_{\chi\chi}^{}\,$,\,
we can translate this limit into a constraint on our GDM parameter space.
We present this constraint in the $\,\MX\!-\!\sqrt{|\xih\xis|}\,$ plane,
as shown in Fig.\,\ref{fig:6}.
Since the invisible width $\,\Gamma(\phi\to\X\X)$\, is proportional to
$\,\left|\!\sqrt{|\xih\xis |}\,\right|^4$,\,
the constraint on $\,\sqrt{|\xih\xis |}$\, is insensitive to either $\,\MX\,$
or the kinetic rescaling factor $\,\zeta\,$ for $\,\phi ZZ\,$ vertex
(given that $\,\xih\,$ itself obeys the LHC bound).
It is also rather insensitive to the experimental limit on the
invisible decay branching fraction around
$\,\text{BR}_{\text{inv}}^{}=\OO(0.1)$.\,
Since the same cubic vertex $\,\phi\X\X\,$ determines both the Higgs invisible decays
and the GDM thermal relic abundance,
we find that the LHC bound on Higgs invisible decays puts a nontrivial constraint
on the required coupling $\,\sqrt{|\xih\xis |}\,$
for generating the observed thermal relic density.
As shown in Fig.\,\ref{fig:6}, we deduce that the GDM with mass
$\,\MX < 48\,$GeV\, is excluded at $95\%$\,C.L.

\vspace*{1mm}

The GDM effective interaction to light fermions in Eq.\,(\ref{eq:FVqqXX})
initiates the $\,\X\X\,$ production via quark annihilation at hadron colliders.
With a mono-jet, photon and $W/Z$ radiation from the initial state quarks $qq'$,\,
the $\,\slashed{E}_T^{}\,$ may be observed. This type of processes has been extensively
studied in literature \cite{ffXX}.
Given the null result, we can infer an upper bound on
$\,|\xih\xis|\,$ as function of the GDM mass $\,\MX\,$.\,
In the high energy regime, $\,q^2\gg m_\phi^2$,\,
the interaction (\ref{eq:FVqqXX}) amounts to an effective operator
$\,\X\X\bar{f}f$.\,
Constraints on the cutoff scale of various effective operators were derived
from combining the results of different initial states measured by ATLAS and CMS
at LHC\,(7\,TeV) \cite{ZBW}. For scalar type operators in our model,
lower bound on $\,\Mp/\!\sqrt{|\xih\xis|}$\, is around $\,\OO(10)$GeV.
The improvement at the LHC\,(8\,TeV) \cite{CMS-8} and the
sensitivity estimated for the LHC\,(14\,TeV) search are fairly mild \cite{ffXX}.
We find that these bounds are quite weak as compared to the constraint from
Higgs invisible decays in Fig.\,\ref{fig:6}.
The case is further studied for the high luminosity LHC and future $pp$ colliders\,\cite{Zhou:2013raa},
but the limit is improved by no more than a factor of 10.
Thus, it appears uneasy to probe the effective interactions
between the GDM and light fermions at hadron colliders.

\vspace*{1mm}

The GDM may be produced from gluon fusions as well, via loop-induced effective operator
$\,\chi_s^2G^{a\mu\nu}G^a_{\mu\nu}$\, in Eq.\,(\ref{eq:FVAAXX}).
As we learn from the SM Higgs production, the suppression from one-loop factor can be compensated
by the large gluon parton distribution function in high energy $\,pp\,$ collisions.
So the gluon fusions provide the most significant production at the LHC\,(14TeV).
However, in comparison with the Higgs production, the loop-factor $\,\mathcal{C}_g\,$
in Eq.\,(\ref{eq:loopfactors}) for the GDM production is energy-dependent
and diminishes for $\,\sqrt{s}\gg m_f^{}\,$.\,
Thus, the production cross section in high energy $pp$ collisions becomes much smaller than
what is expected for the SM Higgs production.
The gluon fusion production of DM as induced by the top-DM effective operator
has been analyzed by using mono-jet searches at the LHC \cite{top-DM}.
This greatly improves the sensitivity over the standard search based on
light fermion effective operators. But, due to the loop-factor suppression,
the lower bound on cutoff scale is around 100\,GeV,
which is still weaker than that derived from generating the thermal relic abundance
by the GDM.

\begin{figure}[t]
\centering%
\includegraphics[height=9cm,width=12.5cm]{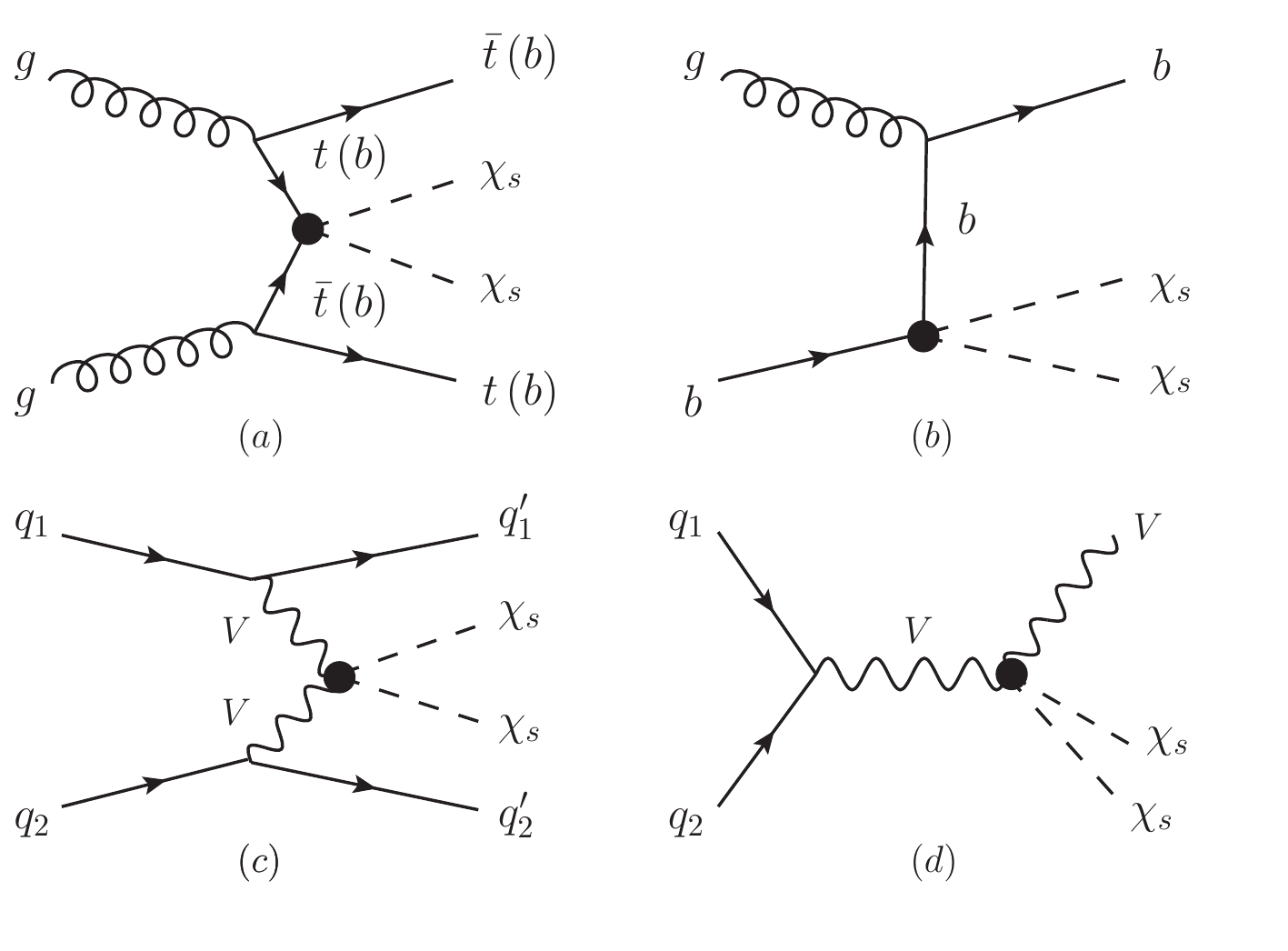}
\vspace*{-2mm}
\caption{GDM production processes at hadron colliders.
Plot-(a): top pair (bottom pair) associated production of GDM particles.
Plot-(b): single bottom associated production of GDM, where the bottom can be
either $b$ or $\bar{b}$\,.\,
Plot-(c): GDM production via vector boson fusion ($V=W,Z$).
Plot-(d): vector boson associated production of GDM. }
\label{fig:7}
\vspace*{3mm}
\end{figure}

In our model, the GDM has much larger couplings to heavy particles. 
Thus, we can effectively produce the GDM pair through its interactions with
the third generation quarks $(t,\,b)$ or the vector bosons $(W,\,Z)$.\, 
Fig.\,\ref{fig:7}(a) presents the top pair (bottom pair) associated production of GDM particles.
Fig.\,\ref{fig:7}(b) shows the single bottom associated production of GDM, where the bottom can be
either $b$ or $\bar{b}$\,.\,
The black dots denote the effective interactions (\ref{eq:FVqqXX}).
In the high energy regime $\,q^2\gg m_\phi^2$\,,\,
the propagator suppression for the dominant Higgs exchange diagram is compensated by energy enhancement
in the $\,\X\!-\!\X\!-\!\phi\,$ vertex, and the effective interactions becomes contact.
The top pair (bottom pair) associated DM production can effectively probe scalar-type interactions between
the DM and quarks\,\cite{Lin:2013sca}.
This is a typical feature of our GDM in the present model.
Recently, CMS presented the analysis for dirac DM in di-leptonic top decay channel \cite{CMStopXX}. 
The latest ATLAS search\,\cite{ATLASbtop} analyzed the DM production in the same processes of 
Fig.\,\ref{fig:7}(a)-(b) by using effective contact operators.
They found that the top associated DM production has higher sensitivity than the bottom associated DM 
production for scalar type operator, where the bottom associated production has larger phase space 
for the final state, but not enough to compensate the suppression effect of bottom Yukawa coupling 
(relative to the top Yukawa coupling).  The lower bound on the cutoff scale for the scalar DM, 
is around $\,\mathcal{O}(10)\,$GeV, which is generally weaker than the Dirac DM. 
So far, this limit is still too weak to constrain the GDM prediction.
Given the heaviness of top quark, we expect a significant improvement of sensitivity in this channel
at future circular $pp$ colliders ($50-100$\,TeV) \cite{FCC}.

\vspace*{1.5mm}

The interactions of DM with weak gauge bosons can initiate the DM production via 
vector boson fusion (VBF) and vector boson associated production, as shown in  
Fig.\,\ref{fig:7}(c) and (d), respectively. 
For the DM production via VBF,  analysis was done for the LHC\,(14\,TeV) in the context of SUSY models\,\cite{Delannoy:2013dla}. With the effective operator approach, Ref.\,\cite{XXVVsearch2} 
recasted the CMS search for invisible Higgs decays in the VBF production and converted the CMS results 
into the bound on the DM mass.
The associated production contributes to mono-$W/Z$ signals. 
Ref.\,\cite{XXVVsearch1} used the ATLAS analysis of $\,W+\slashed{E}_T^{}\,$  
to derive constraints on the DM-vector-boson effective operator. 
But, these studies focused on the gauge invariant operator $\,\chi_s^2 F^{\mu\nu a}_iF^a_{\mu\nu i}$,\, 
which could be generated only at one-loop in our GDM model. 
We note that under the limit $\,q^2\gg m_\phi^2$\,,\, 
our tree-level effective interaction (\ref{eq:FVVVXX}) will induce a contact
operator $\,\chi_s^2V_\mu^{} V^\mu$\,,\, which is not yet studied.
It is encouraging to perform systematical Monte Carlo simulations for the $\,\chi_s^2V_\mu V^\mu$\, 
type operator at the upcoming LHC runs with $13-14$\,TeV collision energy.  
Further studies at the future high energy $pp$ colliders\,($50-100$\,TeV)\,\cite{FCC}
should effectively probe the heavier mass range of the GDM.
This is fully beyond the current scope and will be considered elsewhere.

\section{\hspace*{-3.5mm} Conclusions}
\label{sec:5}
\vspace*{2mm}

All the astrophysical and cosmological evidences of dark matter (DM) so far have
only demonstrated the role of its gravitational interactions.
An intriguing possibility is that the DM communicates with our visible world
via gravitation only.  In this work, we presented a minimal construction of such
a gravitational dark matter (GDM), where a scalar GDM particle \,$\X$\, couples to the
SM through the unique dimension-4 operator \eqref{eq:NMC} which
contains the fields $\,\XX$\, and Ricci curvature $\,{\cal R}\,$.\,

\vspace*{1mm}

In Section\,\ref{sec:2}, we formulated this minimal GDM in both Jordan frame and Einstein frame.
The GDM $\,\X$\, is a real singlet scalar and odd under the $\,\mathbb{Z}_2^{}$\, symmetry,
which may serve as a WIMP DM candidate.
In Jordan frame, both the dark matter particle $\,\X\,$ and the SM Higgs boson $\,\phi$\,
have gravitational interactions \eqref{SRNMCJF} with nonminimal couplings
$\,\xis\,$ and $\,\xih\,$,\, respectively.
Due to the graviton-exchange and the graviton-Higgs kinetic mixing, the interactions
between the dark matter $\X$ and SM particles will be enhanced by the coupling product
$\,|\xis\xih|\gg 1\,$,\, besides the suppression factor $\,(\vew^2,E^2)/\Mp^2\,$.\,
In Einstein frame, these effective interactions become manifest,
as shown in Eqs.\,\eqref{SRNMCEFb} and \eqref{SRNMCEFf}.
Our model only invokes three key parameters in the DM phenomenology:
the GDM mass $\,\MX\,$ and the nonminimal couplings $(\xis,\, \xih)$.\,
For convenience of physical analysis,
we derived all relevant Feynman vertices for the GDM in Einstein frame.
We also derived the perturbative unitarity constraints
on the new couplings $\,(\xis,\, \xih)$,\,
and identified the valid perturbative parameter space in Fig.\,\ref{fig:3},
which justifies our leading order analysis for the GDM.

\vspace*{1mm}

In Section\,\ref{sec:3}, we systematically analyzed the GDM thermal relic density.
For the viable parameter space, we found that the Higgs-exchange contributions dominate the
dark matter annihilation cross sections, where only the dark matter mass $\,\MX\,$ and
coupling product $\,\xih\xis\,$ are relevant. Since the leading order interactions between GDM
and SM fields are proportional to the corresponding SM particle masses,
the DM annihilations into heavy modes dominate in the large $\,\MX\,$ range.
In Fig.\,\ref{fig:3}, the compatibility of the predicted GDM thermal relic abundance
with Planck data was demonstrated
in the $\,\MX-\sqrt{|\xih\xis|}$\, plane for a wide range of $\,\X\,$ mass.
The red solid curve of Fig.\,\ref{fig:3} corresponds to the central value
of $\,\Omega_{\chi 0}^{}h^2\simeq 0.12\,$.\,
Since the GDM cross sections are proportional to $\,(\!\sqrt{|\xih\xis|})^4$,\,
the predicted parameter space of $\,\sqrt{|\xih\xis|}\,$ in Fig.\,\ref{fig:3} has little
sensitivity to the experimental uncertainty of $\,\Omega_{\chi 0}^{}h^2$.\,

\vspace*{1mm}

In Section\,\ref{sec:4}, we further studied possible direct and indirect detections of GDM,
as well as discussing its collider searches.
Direct detection of GDM relies on its effective interactions with light fermions.
In comparison with the GDM annihilation cross sections, the GDM-nucleon scattering is largely
suppressed for the small momentum exchange. As shown in Fig.\,\ref{fig:4},
the required range of $\,\sqrt{|\xih\xis|}\,$ for accommodating the thermal relic abundance
predicts signals to be even lower than the general neutrino background, so it is
out of reach of the current direct detection technique.
For indirect detections, we mainly studied constraints from the observation
of gamma ray spectrum which is most promising for searching the GDM.
The line spectrum arises from direct annihilations of dark matter into $\gamma$\,'s.
These operators are generated at one-loop level for the GDM.
The diffuse continuum spectrum reflects the secondary photons produced from
primary dark matter annihilations into massive gauge bosons, quarks or leptons.
We summarized the constraints for these processes in the GDM parameter space,
as shown in Fig.\,\ref{fig:5}.
In contrast to the direct detection, we found that gamma ray searches are promising,
and have higher sensitivity to the heavier GDM particles.
For $\,\MX\gtrsim \OO(100)$GeV, the prediction of our model
is within the reach of future gamma ray searches of diffused spectrum.
For collider searches, we first studied the constraint from measuring Higgs invisible decays at the LHC.
We derived a nontrivial upper bound on $\,\sqrt{|\xih\xis|}$\, for low $\MX$ region,
which excludes the GDM with mass $\,\MX < 48\,$GeV at 95\%\,C.L.
Finally, we discussed the searches of $\,\X\,$ at hadron colliders with different production channels. 
The GDM pair production in association with top (bottom) pair, or with single bottom-jet or 
mono-$W/Z$, or from the vector-boson-fusions (Fig.\,\ref{fig:7}) can be probed at 
the upcoming LHC runs and the future high energy $pp$ colliders \cite{FCC}.


\vspace*{8mm}
\noindent
{\bf\large Appendix:}
\vspace*{1mm}

\appendix
\section{\hspace*{-3mm} Formulas for Radiative Loop Factors}
\label{app:loopfactor}
\vspace*{2mm}

In this Appendix, we summarize the exact expressions of the loop factors\,\cite{Gunion:1989we}
for effective interactions between the dark matter and gauge bosons.
The loop factors $(\mathcal{C}_g,\, \mathcal{C}_\gamma )$ in Eq.\,\eqref{eq:loopfactors} contain
only a single mass-ratio $\,\tau_j^{}=E^2/4m_j^2$\, via functions $(A_V^{},\,A_F^{})$,
where \,$E$\, denotes the center of mass energy.
The functions $(A_V^{},\,A_F^{})$ are defined as follows,
\begin{eqnarray}
  A_V(\tau)&=&\frac{1}{8\tau^2}[3\tau+2\tau^2-3(1-2\tau)f(\tau)],
  \\[2mm]
  A_F(\tau)&=&\frac{3}{2\tau^2}[\tau-(1-\tau)f(\tau)],
  \\[2mm]
  f(\tau)&\equiv&\left\{\begin{array}{ll}
  \arcsin^2\sqrt\tau\,, &~~~~ \tau\leqq 1\,,
  \\[2mm]
  -\dis\frac{1}{4}
  \left[\ln\frac{1+\sqrt{1-\tau^{-1}}}{1-\sqrt{1-\tau^{-1}}}-i\pi\right]^2,
  &~~~~  \tau>1 \,.
  \end{array}\right.
\end{eqnarray}

Since $Z$ is massive, the loop factor $\,\mathcal{C}_{\gamma z}\,$ in Eq.\,\eqref{eq:loopfactors}
involves another mass ratio $\,\eta_j^{}=m_Z^2/4m_j^2$.\,
The functions $(B_V^{},\,B_F^{})$ are defined as follows,

\begin{eqnarray}
  B_V^{}(\tau,\eta ) &\,=\,& -t_W^{-1}\left[4(3-t_W^2)I_2^{}(\tau,\eta)
  +\left((1+2\tau)t_W^2-(5+2\tau)\right)I_1^{}(\tau,\eta )\right] ,
\\[2mm]
  B_F(\tau,\eta)&=&
  N_C^{}\frac{-2Q_f^{}(T_f^{3L}-2Q_f s_W^2)}{s_W^{}c_W^{}}[I_1^{}(\tau,\eta)-I_2^{}(\tau,\eta)],
\\[2mm]
  I_1^{}(\tau,\eta)&=&\frac{\tau^{-1}\eta^{-1}}{\,2(\tau^{-1}\!-\!\eta^{-1})\,}
  +\frac{\tau^{-2}\eta^{-2}}{\,2(\tau^{-1}\!-\!\eta^{-1})^2\,}\left[f(\tau)\!-\! f(\eta )\right]
  +\frac{\tau^{-2}\eta^{-1}}{(\tau^{-1}\!-\!\eta^{-1})^2}\left[g(\tau)\!-\! g(\eta )\right]\!,
  \hspace*{10mm}
\\[2mm]
  I_2^{}(\tau,\eta ) &=&
  -\frac{\tau^{-1}\eta^{-1}}{2(\tau^{-1}\!-\!\eta^{-1})}\left[f(\tau)-f(\eta)\right],
\\[2mm]
  g(\tau) & = & \left\{\begin{array}{ll}
  \sqrt{\tau^{-1}\!-1\,}\arcsin\sqrt\tau\,, &~~~~ \tau\leqq 1\,,
  \\[3mm]
  \frac{1}{2}\sqrt{1\!-\tau^{-1}\,}\left[\log\frac{1+\sqrt{1-\tau^{-1}}}{1-\sqrt{1-\tau^{-1}}}-i\pi\right]\!,
  &~~~~ \tau >1\,.
  \end{array}\right.
\end{eqnarray}
where we have defined
$\,(s_W^{},\,c_W^{})\equiv (\sin\theta_W^{},\,\cos\theta_W^{})$,\,
and $\,t_W^{} \equiv \tan\theta_W^{}$,\, with $\,\theta_W^{}\,$ denoting the weak mixing angle.
Also, $\,N_C^{}=3\,(1)\,$ corresponds to the color factor of quarks (leptons).

\vspace*{4mm}
\section{\hspace*{-3mm} Threshold and Resonance Effects in Thermal Relic Density Analysis}
\label{app:TRDdetail}
\vspace*{2mm}

In this Appendix, we present the calculation of thermal relic density by including
the threshold and resonance effects. Given the mass-spectrum of SM particles,
we note that the two effects take place in difference mass-ranges and can be treated separately.

\vspace*{1mm}

\begin{figure}[t]
\centering%
\includegraphics[height=6.95cm,width=7.6cm]{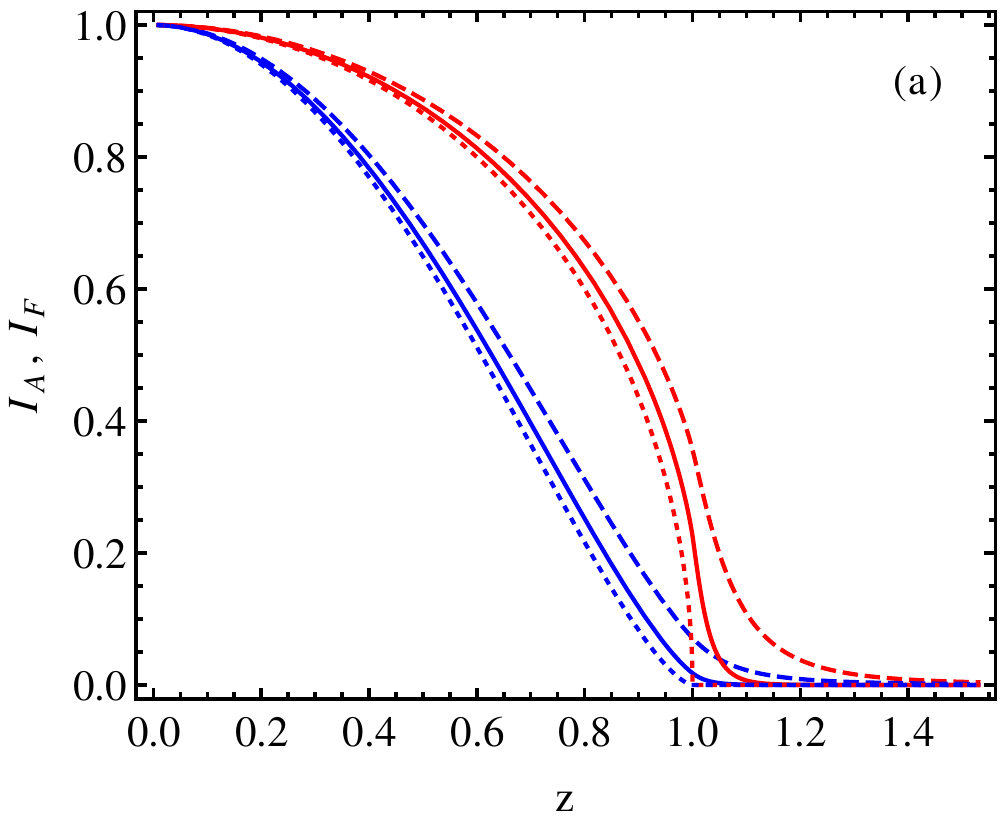}
\includegraphics[height=6.87cm,width=7.6cm]{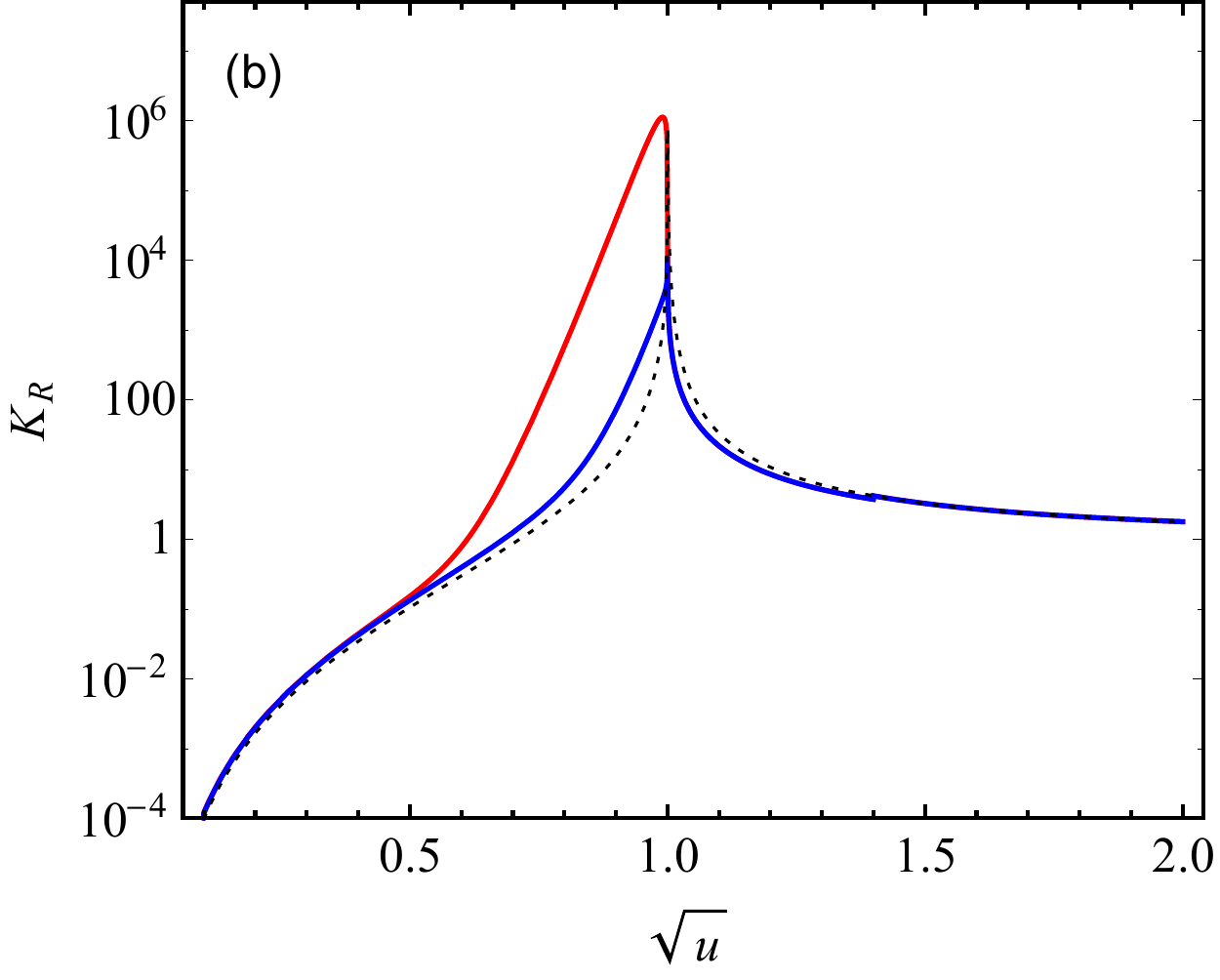}
\vspace*{-2mm}
\caption{Plot-(a): Integrals $\,I_{\text{A}}^{}\,$ and $\,I_{\text{F}}^{}\,$ as functions
of variable $\,z\,$.\,
The red and blue curves correspond to bosonic and fermionic final states, respectively.
The (dotted,\,solid,\,dashed) curves denote inputs of $\,x=(\infty,\, 25,\, 10)$,\,
respectively. Plot-(b): The  thermal averaged integral
$\,K_R^{}\,$ as a function of mass ratio $\,\sqrt{u}=2\MX/m_\phi^{}\,$,\,
with $\,\epsilon$\, given by  $\,\epsilon = \Gamma_\phi^{}/m_\phi^{}$.\,
The dotted curve denotes $\,x=\infty$,\,
and the (red, blue) solid curves represent $\,x=25$\, with the sample inputs
$\,\sqrt{|\xih\xis|}=(10^{15},\,10^{16})$\,,\, respectively.}
\label{fig:8}
\end{figure}

We first consider the threshold effect.
For a generic DM annihilation process $\,\X\X \to f_j^{} f_j^{}\,$,\,
the zero-temperature cross section can be parameterized as
\begin{eqnarray}
\label{eq:sigmaAVpara}
(\sigma_A^{} v) \,=\, (a+b v^2)v_j^n\, ,
\end{eqnarray}
where $\,v\,$ is relative velocity of two dark matter particles,
and $\,v_j^{}\,$ is the final state velocity from phase space integration.
The parameters $\,a\,$ and $\,b\,$ represent $s$-wave and $p$-wave contributions, respectively.
For scalar dark matter, we have $\,n=1\,(3)\,$ for bosonic (fermionic) final states.
Under non-relativistic approximation for the DM, we derive
\begin{eqnarray}
\label{eq:v2z}
v_j^{} ~=~ z\,\sqrt{\frac{v^2}{4}+\mu_+^2\,}\, ,
\end{eqnarray}
where $\,z\equiv m_j^{}/\MX$\, and $\,\mu_+^2\equiv (1-z^2)/z^2$.
Around freeze-out temperature $\,T_{\text{f}}^{}$\,,\, the DM particle is non-relativistic and
the thermal average cross section can be derived by integrating over the relative velocity,
\begin{eqnarray}
\label{eq:sigmaAVapprox}
\langle\sigma_A^{}v\rangle \,=\,
\frac{x^{3/2}}{\,2\sqrt\pi\,}
\int_0^\infty\! dv\,v^2 e^{-\frac{x}{4}v^2}_{} (\sigma_A^{}v)\, + \mathcal{O}(x^{-1},v^2)\, ,
\end{eqnarray}
where $\,x\equiv M_\chi/T\,$. For cold dark matter, we have $\,\xfo\gg 1\,$.\,
Substituting parametrization (\ref{eq:sigmaAVpara}) into (\ref{eq:sigmaAVapprox}),
we deduce the approximate thermal averaged cross section for $\,\MX\geqq m_j^{}$\,,
\begin{eqnarray}
\label{eq:TACSa}
\langle\sigma_A^{}v\rangle_{\text{A}}^{} \,=\, \frac{\,2z^n}{\sqrt{\pi}\,}\int_0^\infty
\!dt\,e^{-t}_{}\!\(\!a+\frac{4bt}{x}\)\!\sqrt{t\left(\frac{t}{x}+\mu_+^2\right)^{\!\!n}\,} \, .
\end{eqnarray}
In the kinematically forbidden case of $\,\MX < m_j^{}\,$,\,
the nonzero velocity $\,v\,$ in  \eqref{eq:sigmaAVpara} 
could make the annihilation viable,
which sets a lower bound of $\,v\,$ in the thermal integration.
For $\,\MX <m_j^{}$,\, we define $\,\mu_-^2\equiv-\mu_+^2>0$\, and $\,v^2>4\mu_-^2$\,.\,
Imposing $\,v>2\mu_-\,$ in (\ref{eq:sigmaAVapprox}) and making change of variables,
we derive the cross section for $\,\MX < m_j^{}\,$,
\begin{eqnarray}
\label{eq:TACSf}
\langle\sigma_A^{}v\rangle_{\text{F}}^{} \,=\,
e^{-x\mu_-^2}\frac{\,\,2z^n}{\sqrt{\pi}\,}\int_0^\infty\! dt\,e^{-t}\!
\left[\(a\!+\!4b\mu_-^2\)\!+\!\frac{\,4bt\,}{x}\right]\!
\sqrt{t\left(\frac{t}{x}\right)^{n-1}\!\!\left(\frac{t}{x}+\mu_-^2\right)\,}\, .
\hspace*{6mm}
\end{eqnarray}
Note that Eqs.\,(\ref{eq:TACSa}) and (\ref{eq:TACSf}) agree at the threshold,
i.e., $\,z\simeq 1\,$ and $\,\mu_+^{}\simeq \mu_-^{}\simeq 0$\,.\,
Since the gravity-induced interaction only generates $s$-wave contribution at leading order,
we will set $\,b=0\,$ afterwards and focus on the $\,a\,$ term in Eq.\,\eqref{eq:sigmaAVpara}
for the following discussion.
For each channel, we may infer $a$ from cross section in (\ref{eq:sAvSMT0}) divided by
$\,v_j^n\,$ at $\,v=0\,$,\, i.e., $(1-z^2)^{n/2}$.
Then, the $s$-wave thermal averaged cross section can be parameterized as
\begin{eqnarray}
\langle\sigma_A^{}v\rangle_{\text{A}}^{} \,\equiv\, a I_{\text{A}}^{}(z,x,n)\,,
\quad\quad
\langle\sigma_A^{}v\rangle_{\text{F}} \,\equiv\, a I_{\text{F}}^{}(z,x,n)\,.
\end{eqnarray}
In Fig.\,\ref{fig:8}(a), we depict $\,(I_{\text{A}}^{},\, I_{\text{F}}^{})$\,
as functions of $\,z\,$ for different $\,x\,$ and $\,n\,$.\,
Red and blue curves denote the cases with bosonic and fermionic final states, respectively.
The (dotted,\,solid,\,dashed) curves correspond to $\,x=(\infty,\, 25,\, 10)$.\,
It is clear that the higher temperature (i.e., smaller $x$) leads to more enhanced thermal integral
from the threshold effect. Comparing the red and blues curves, we see that the annihilation cross section
with bosonic final states is more enhanced.

\vspace*{1mm}

Next, we consider the resonance effect, which is important around $\,\MX \sim m_\phi^{}/2$.\,
From (\ref{eq:sAvSMT0}), we see a common factor from Higgs-exchange for both $\,f\bar{f}\,$
and $\,VV\,$ final states. Taking into account the finite temperature effect, we have
$\,s\simeq 4M_\chi^2/(1\!-\!v^2/4)$\, and the expressions are modified in following way,
\begin{eqnarray}
\label{eq:p-factor}
\frac{(4M_\chi^2)^2}{\,(4M_\chi^2-m_\phi^2)^2+m_\phi^2\Gamma_\phi^2\,}
~\rightarrow~
\frac{u^2/(1-v^2/4)^2}{\,(1-u/(1-v^2/4))^2+\epsilon^2\,}\, ,
\end{eqnarray}
where $\,u\equiv 4M_\chi^2/m_\phi^2$\, and $\,\epsilon\equiv \Gamma_\phi^{}/m_\phi^{}\,$.\,
For non-relativistic GDM, the thermal averaged integration over the propagator factor
\eqref{eq:p-factor} defines the following function,
\begin{eqnarray}
K_R^{}(x,u,\epsilon) \,=\, \frac{\,x^{3/2\,}}{\,2\sqrt\pi\,}\int_0^\infty \!dv\,v^2 e^{-\frac{x}{4}v^2}_{}\frac{u^2/(1-v^2/4)^2}{\,\left[1-u/(1-v^2/4)\right]^2+\epsilon^2\,}\, .
\end{eqnarray}
As shown in Ref.\,\cite{Griest:1990kh}, for narrow resonance like the SM Higgs boson,
i.e., $\,\epsilon\sim 10^{-5}$, any expansion over $v^2$ may yield considerable error
around the resonance pole. Thus, we will perform thermal integration numerically for computing  $\,\langle\sigma_A^{}v\rangle$\,.\, For $\,\MX >m_\phi^{}/2\,$,\, i.e.,
$\,u>1$,\, the width effect quickly becomes subdominant and negligible.
In the light mass range, $\,\MX\lesssim m_\phi^{}/2$,\,
the cross section is more enhanced due to finite temperature integration.
Also, the Higgs invisible decay starts to open and the width depends on
$\MX$ and $\,\xih\xis\,$.\,
Fig.\,\ref{fig:8}(b) depicts $\,K_R^{}\,$ as a function of $\,\sqrt{u}=2\MX/m_\phi^{}\,$ with
$\,\epsilon=\Gamma_\phi^{}/m_\phi^{}\,$.\,
For illustration, we choose $\,\sqrt{|\xih\xis|}=10^{15},\,10^{16}$\,
as two benchmarks. The dotted curve corresponds to $\,x=\infty$\,,\,
where little difference can be seen for the two cases.
The (red, blue) solid curves represent $\,\sqrt{|\xih\xis|}=(10^{15},\,10^{16})$\,
at $\,x=25$\,,\, respectively.
For the case of $\,\sqrt{|\xih\xis|}=10^{15}$\,,\, we see significant resonance enhancement for
$\,\sqrt{u}\lesssim 1\,$,\, as compared with the zero-temperature estimate.
The other case of $\,\sqrt{|\xih\xis|}=10^{16}$\,
corresponds to a much larger Higgs width and thus the ratio
$\,\epsilon \,(= \Gamma_\phi^{}/m_\phi^{})\,$
since $\,\Gamma [\phi\to\X\X]\propto |\xih\xis|^2\,$ [cf.\ Eq.\,\eqref{eq:width-H-XX}].
As shown in Fig.\,\ref{fig:8}(b), the resonance effect is much smaller in this case.

\acknowledgments    
\vspace*{-2mm}
We thank Haipeng An, Antonio Boveia, James Cline, Bob Holdom, C.\ S.\ Lam, Tim Tait, Yue Zhang, 
Ning Zhou, and Kathryn Zurek for discussions and correspondences.
We thank Xavier Calmet for collaboration on an earlier version of GDM
and for related discussions on the manuscript.
This research was supported by National NSF of China (under grants 11275101, 11135003)
and National Basic Research Program (under grant 2010CB833000).



\end{document}